\documentclass[journal]{IEEEtran}
\usepackage{amsmath,amsfonts}
\usepackage{amssymb}
\usepackage{algorithm}
\usepackage[indLines=true,noEnd=false]{algpseudocodex}
\usepackage{array}
\usepackage[caption=false,font=normalsize,labelfont=sf,textfont=sf]{subfig}
\usepackage{textcomp}
\usepackage{threeparttable}
\usepackage{booktabs}
\usepackage{graphicx}
\usepackage{float}
\usepackage{stfloats}
\usepackage{url}
\usepackage{verbatim}
\usepackage{cite}
\usepackage{bm}
\usepackage
[colorlinks=true, 
linkcolor=blue, 
citecolor=blue, 
urlcolor=cyan, 
bookmarksopen=true]{hyperref}
\usepackage{xcolor}

\begin{document}

\title{Hybrid-Field Sparse Channel Representation and Recovery for XL-RIS-Assisted mmWave MIMO Systems}
\author{
Wenkai Liu, 
Nan Ma,~\IEEEmembership{Member,~IEEE,} 
Jianqiao Chen,
Hongtao Zhang,~\IEEEmembership{Senior Member,~IEEE,} 

and 
Ping Zhang,~\IEEEmembership{Fellow,~IEEE} 
\thanks{
Wenkai Liu, Nan Ma, Hongtao Zhang and Ping Zhang are with Beijing University of Posts and Telecommunications, Beijing 100876, China. 
Nan Ma and Ping Zhang are also with the Department of Broadband Communication, Peng Cheng Laboratory, Shenzhen 518055, China 
(e-mail: liuwenkai@bupt.edu.cn; manan@bupt.edu.cn; htzhang@bupt.edu.cn; pzhang@bupt.edu.cn).(Corresponding author: Nan Ma.)

Jianqiao Chen is with ZGC Institute of Ubiquitous-X Innovation and Applications, Beijing 100876, China (e-mail: jqchen1988@163.com).
}}
\markboth{IEEE Transactions on Wireless Communications,~Vol.~XX, No.~XX, 2026}%
{Shell \MakeLowercase{\textit{et al.}}: A Sample Article Using IEEEtran.cls for IEEE Journals}

\maketitle
\begin{abstract}
Extremely large-scale reconfigurable intelligent surface (XL-RIS)-assisted communication is regarded as a key enabling technology for future 6G networks. 
However, hybrid-field channel estimation for XL-RIS-assisted systems is challenging due to the high-dimensional cascaded channel and the coexistence of far-field and near-field propagation.
In this case, traditional full-dimensional sparse recovery methods require a large cascaded dictionary and suffer from severe computational and storage burdens.
To address these challenges, we develop a double-timescale channel estimation framework that decouples sparse dictionary representation and recovery.
Then, by exploiting the quasi-static property of the channel at the base station (BS) and RIS side, we propose a Dirichlet kernel-based off-grid dictionary compression (DK-ODC) scheme for sparse representation, which reduces the dimension of the corresponding dictionary as well as mitigates BS-side angular off-grid error.
Furthermore, for the dynamic channel at the user equipment (UE) and RIS side, we propose a subspace-aware incremental variational Bayesian learning (SI-VBL) algorithm, which enables incremental learning of sparse channels by exploiting the identified low-dimensional subspace and pruning threshold. 
Analysis and simulation results confirm that the proposed framework avoids full-dimensional Bayesian recovery and achieves a favorable tradeoff among estimation accuracy, computational complexity, and storage overhead.
\end{abstract}

\begin{IEEEkeywords}
XL-RIS, sparse channel representation, channel estimation.
\end{IEEEkeywords}

\section{Introduction}
\IEEEPARstart{T}{he} reconfigurable intelligent surface (RIS) has emerged as a promising technology for millimeter-wave (mmWave) and terahertz (THz) multiple-input multiple-output (MIMO) systems \cite{tang2021wireless-1}, \cite{najafi2021physics-2}, \cite{pei2021ris-3},
\cite{yuan2021reconfigurable-4},
\cite{zheng2022survey-5}. 
The path loss of the RIS reflection link depends on the product of the path lengths from the UE to the RIS (UE-RIS) and from the RIS to the BS (RIS-BS) rather than their sum \cite{wei2022codebook}, leading to signal degradation in conventional RIS systems over long distances.
Extremely large-scale RIS (XL-RIS) improves passive beamforming gain by increasing the number of elements, compensating for path loss and enhancing coverage quality in weak signal areas \cite{6G_1}. 
Thus, RIS is evolving towards XL-RIS, which is expected to play a key role in future 6G communications \cite{6G_1}, \cite{6G_2}.
To fully reap these benefits, it is fundamental to acquire accurate CSI, which will be further used in beamforming and other signal processing tasks \cite{wymeersch2020radio-6},\cite{ H. Zhang}, \cite{lin2021tensor-8}, \cite{alexandropoulos2022localization-9}. 
However, given that MIMO arrays and the RIS collectively comprise hundreds of elements, the dimension of the cascaded channel matrix increases significantly. 
Estimating the cascaded channel with a passive XL-RIS leads to huge pilot overhead and computational complexity \cite{pilot overhead and complexity}.

\subsection{Prior Work}
Compressed sensing (CS) leverages the inherent sparsity of wireless channels, emerging as a promising paradigm for RIS-assisted channel estimation. 
By formulating the estimation task as a sparse signal recovery problem, these methods can reduce the pilot overhead.
Several CS-based channel estimation algorithms have been developed for RIS-assisted systems, including orthogonal matching pursuit (OMP)-based methods \cite{OMP}, atomic norm minimization \cite{ANM}, and approximate message passing (AMP)-based approaches \cite{AMP}.
The authors in \cite{SBL} investigated the Kronecker-structured sparse reconstruction task under the sparse Bayesian learning (SBL) framework, where two algorithms based on alternating minimization and singular value decomposition are developed.
However, CS-based methods suffer from the unavoidable off-grid error, since the dictionary matrix is constructed by quantizing the direction vectors at the specific resolution \cite{off-grid}.
From the perspective of parameter estimation, the CS-based parameter recovery problem was established to obtain holographic RIS-assisted THz massive MIMO cascaded channel \cite{THz}.
In \cite{canshuguji}, the authors employed a CS method to recover angles of departure (AODs), angles of arrival (AoAs), and channel gains used to construct cascaded channel.
These methods disregard quasi-static RIS-BS channel, whose stability stems from fixed deployment, thereby simplifying the estimation process and shifting the estimation focus to the dynamic UE-RIS link. 
This approach reduces both pilot overhead and complexity \cite{quasi-static}.
Thus, the authors in \cite{Two-Timescale} proposed the two-time scale channel estimation framework and estimated the RIS-BS channel using a double-link pilot transmission scheme.
In \cite{SA-LS}, the authors divided the estimation process into slow-time parameter estimation and fast-time parameter estimation patterns, and proposed a subspace-aware least squares (SA-LS) method to estimate the channel gain.

The above scheme effectively solves the problem of far-field channel estimation for narrow band systems utilizing sparsity in the angular domain. 
As the aperture of the XL-RIS increases, the range of its corresponding Fresnel region (i.e., the radiated near-field region of the antenna array) also expands further. 
Under this mode, the array response vector of the channel is dependent on both angle and distance, making conventional angular domain dictionary ineffective.
To address this, a polar-domain sparsity framework was established for extremely
large-scale multiple-input multiple-output (XL-MIMO) \cite{polar}, followed by tailored algorithms such as polar-domain OMP \cite{polar} and parametric estimation based on common support properties \cite{near-PCE}.
More recently, SBL-based approaches have been extended to the near-field \cite{near-sbl, near-off-grid-sbl} to enhance estimation precision.
Nevertheless, a critical bottleneck of these polar-domain methods is the prohibitive computational complexity and storage overhead derived from the high-dimensional dictionary, particularly for XL-RIS. 
Although recent attempts, such as the distance-parameterized angular model \cite{distance-Para} and adaptive Joint SBL (JSBL) \cite{JSBL}, seek to reduce dictionary size for XL-MIMO systems, the XL-RIS cascaded channel estimation remains an immediate challenge to be addressed.

Existing research indicates that XL-RIS-assisted systems tend to operate in a hybrid-field channel environment.
However, XL-RIS hybrid-field cascaded channel estimation is more complex, requiring solutions to hybrid-field cascaded channel modeling, near/far-field component separation, and cascaded parameter estimation.
Building upon the solid foundation of far-field and near-field channel estimation, preliminary progress has been achieved in hybrid-field channel estimation for XL-RIS-assisted systems \cite{XL-RIS-EM}, \cite{HSGP}.
For the problem of cascaded hybrid-field channel estimation, an effective solution is to jointly exploit the angular-domain sparsity of far-field channels \cite{OMP} and the polar-domain sparsity of near-field channels \cite{polar}. 
Following this idea, \cite{XL-RIS-EM} achieved joint iterative inference and update of distribution parameters through an expectation-maximization algorithm. 
To improve efficiency, \cite{HSGP} further proposed a dedicated hybrid-field stochastic gradient pursuit (HSGP) algorithm.

In this paper, to compensate for the limited antenna gain at the UE \cite{antenna few}, \cite{user}, the XL-RIS is deployed near the UE. 
This configuration establishes a coupled hybrid-field channel, where the RIS-BS link operates in the far-field while the UE-RIS link exhibits near-field characteristics \cite{3D-D-LAOMP}.
In this case, the hybrid-field channel estimation method mentioned in \cite{XL-RIS-EM}, \cite{HSGP} is not suitable.
In \cite{3D-D-LAOMP}, the authors introduced 3D-multiple measurement vector compressed sensing  (3D-M-CS) and 3D-distributed CS (3D-D-CS) frameworks solved by look-ahead OMP (LAOMP) variants, they achieve on-grid recovery of both angular and polar-domain parameters with reduced pilot overhead.
To address limited visibility regions in XL-RIS systems, the work in \cite{sub-SBL} proposed a channel estimation method based on fast sparse Bayesian learning.
This method employed the shifted common-support property among sub-channels to perform joint estimation across all sub-channels.
In essence, these methods depend on constructing grid-based sparse representation dictionary and employing corresponding sparse recovery algorithms to estimate the channel.
In this context, we still believe that hybrid-field cascade channel estimation needs to be further explored and addressed to overcome the following challenges. 
For sparse channel representation, the excessive dimensionality of overcomplete cascaded dictionary leads to prohibitive storage overhead, especially in XL-RIS cascaded channels where the associated memory footprint can become prohibitive for memory-limited online processing platforms \cite{JSBL}, \cite{Liu_Low}.
Moreover, enforcing discretized physical grids inevitably induces off-grid error, resulting in degraded estimation accuracy \cite{Cao_Efficient}, \cite{Chu2024PDANM}. 
For sparse channel recovery, there is an urgent need to develop sparse recovery algorithms that achieve a balance between computational efficiency and estimation accuracy.
\subsection{Our Contributions}
Therefore, it is imperative that channel estimation strategy for hybrid-field XL-RIS systems be tailored to address these specific issues. 
Our main contributions of this paper are summarized as follows:
\begin{itemize} 
 \item \textit{Hybrid-Field Channel Estimation Problem Formulation:} 
Considering the potential coexistence of far-field and near-field effects in XL-RIS channels, we first simplify the modeling of hybrid-field channel and design a joint sparse representation based on angular and polar domains. Then we explore the inherent space-time sparsity of hybrid-field channels, formulating a channel estimation strategy comprising dictionary compression representation and sparse channel recovery.
\item \textit{Low-Dimensional Sparse Channel Representation for Quasi-Static RIS-BS Link:}
To mitigate the high dimensionality of the sparse representation dictionary as well as off-grid error caused by uniform angular-sampling, we propose a Dirichlet kernel-based off-grid dictionary compression (DK-ODC) scheme. 
Specifically, instead of refining angles through dense discretization, it transforms continuous angle estimation into Dirichlet kernel peak localization within a bounded region, which enables super-resolution angle recovery and compresses the corresponding sparse dictionary by orders of magnitude.

\item \textit{Efficient Sparse Channel Recovery for Dynamic UE-RIS link: }
Based on the above-mentioned dictionary, we propose a subspace-aware incremental variational Bayesian learning (SI-VBL) method for estimating the dynamic UE-RIS channel. 
Specifically, it introduces a correlation-driven subspace screening mechanism to confine Bayesian learning to a low-dimensional active manifold.
Meanwhile, derived from marginal evidence maximization, a likelihood-based pruning rule enables adaptive path selection under highly coherent polar dictionaries. 
Our method retains the robustness of Bayesian learning-based methods while enabling efficient sparse channel recovery via avoidance of large-scale matrix inversion and full-dimensional processing.

\item \textit{Applicability Validation Across Extended Models:}
We validate the applicability of the proposed framework in extended XL-RIS models, including multi-user pilot-contaminated model and UPA-based deployments. 
These extensions examine the robustness of the proposed estimator under pilot reuse and its applicability to higher-dimensional array geometries.
Complexity and storage analyses, together with ULA and UPA simulations, show that the proposed framework achieves a favorable tradeoff among estimation accuracy, computational complexity, and storage overhead.
\end{itemize}

\subsection{Organization}
The remainder of this paper is organized as follows.
In Section \ref{II}, we introduce the system model and problem formulation for XL-RIS hybrid-field. 
In Section \ref{III}, we propose the method of channel estimation based on sparse channel representation and recovery.
In Section \ref{IV}, we present model extensions and computational complexity
analysis. 
Numerical results are provided in Section \ref{V}. 
Finally, we conclude the paper in Section \ref{VI}.

\textit{Notation:} Bold lower-case and upper-case letters denote vectors and matrices, respectively. 
The operators \((\cdot)^T\), \((\cdot)^H\), \((\cdot)^*\), \((\cdot)^{-1}\), and \((\cdot)^\dagger\) denote transpose, Hermitian transpose,
conjugate, inverse, and pseudo-inverse, respectively.
The operators \(\otimes\), \(\odot\), \(\mathrm{vec}(\cdot)\), and \(\mathrm{tr}(\cdot)\) denote Kronecker product, Hadamard product, vectorization, and trace. Finally, \(\mathcal{CN}(\boldsymbol{\mu},\boldsymbol{\Sigma})\) denotes a complex Gaussian distribution.
\section{System model and problem formulation}
\label{II}
As illustrated in Fig. \ref{fig_model}, we consider a mmWave MIMO uplink time-division duplexing (TDD) system where the XL-RIS is deployed near a single-antenna UE to assist the signal transmission between the UE and the BS. 
The BS is equipped with \(M\) antennas, and the XL-RIS consists of \(N\) reflecting elements arranged as a uniform linear array (ULA)\footnote{The extension to the uniform planar array (UPA)-based XL-RIS is discussed in Section~\ref{subsec:UPA}.}.
Without loss of generality, we focus on the scenario where there is no direct link between the BS and the UE. 
In this case, the cascaded channel \textbf{G} among the BS, RIS and UE can be defined as
\begin{equation}
    \mathbf{G} \triangleq \mathbf{G}_{\mathrm{RB}} \cdot \operatorname{diag}(\mathbf{h}_{\mathrm{UR}}) \in \mathbb{C}^{M \times N},
    \label{eq:1}
\end{equation}
where $\mathbf{G}_{\mathrm{RB}} \in \mathbb{C}^{M \times N}$ denotes the channel between the RIS and BS, and $\mathbf{h}_{\mathrm{UR}} \in \mathbb{C}^{N\times1}$ denotes the channel from UE to the RIS.
For the BS to estimate the channel, the UE transmits pilot symbols \(m_t\), 
\(t\in\{1,2,\ldots,T\}\), to the BS over \(T\) time slots\footnote{
In a multi-user scenario, different UEs are assigned pilot sequences for channel estimation. 
With mutually orthogonal pilots, different UEs can be separated after matched filtering, and each cascaded channel \(\mathbf{G}_j\) can be estimated by the single-user model in \eqref{eq:2}. 
When the available pilot length is insufficient, pilot reuse or non-orthogonal assignment causes pilot-contaminated observations, thereby degrading channel estimation accuracy \cite{Jose2011}. 
The multi-user extension with pilot contamination is discussed in Section~\ref{subsec:multi_user_pc}.}.
For simplicity, we assume $m_t = 1$ for all $t$. 
The measurement matrix 
$\mathbf{Y} = [\mathbf{y}_1, \mathbf{y}_2, \ldots, \mathbf{y}_T] \in \mathbb{C}^{M \times T}$ received at the BS from the UE can be expressed as
\begin{equation}
\mathbf{Y} = \mathbf{G}\bm{\Gamma} + \mathbf{E},
\label{eq:2}
\end{equation}
where $\bm{\Gamma} = [\Upsilon_1, \Upsilon_2, \cdots, \Upsilon_T] \in \mathbb{C}^{N \times T}$ and $\mathbf{E} = [\bm{\varepsilon}_1, \bm{\varepsilon}_2, \cdots, \bm{\varepsilon}_T] \in \mathbb{C}^{M \times T}$ denote the phase reflection matrix and the additive white Gaussian noise (AWGN) matrix, respectively.
$\boldsymbol{\Upsilon}_t=[\upsilon_{1,t},\ldots,\upsilon_{N,t}]^T$ denotes the RIS training reflection vector at the $t$-th pilot slot. 
In the simulations, we set $\upsilon_{n,t}\in\{+1/\sqrt{N},-1/\sqrt{N}\}$.
Here, $\bm{\varepsilon}_t \in \mathbb{C}^{M \times 1}$ follows $\mathcal{CN}(\bm{\varepsilon}_t \mid \mathbf{0}, \sigma^2\mathbf{I}_M)$, where $\sigma^2$ denotes the noise variance.
\subsection{Channel Model}
\begin{figure}[!t]
\centering
\includegraphics[width=2.8in]{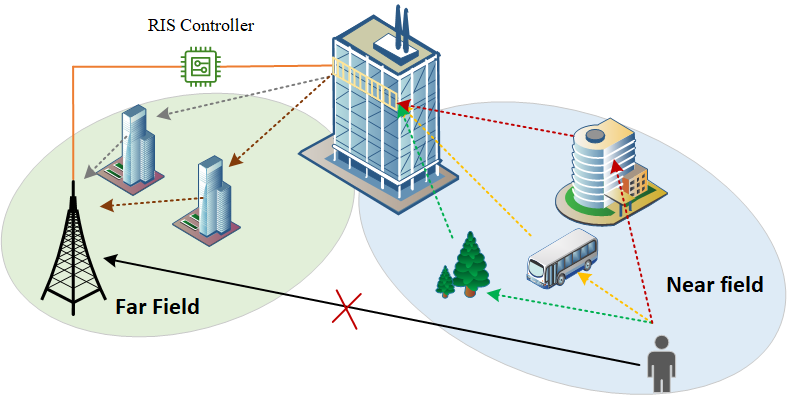}
\caption{
System model.}
\label{fig_model}
\end{figure}

Due to the substantial physical sizes of ULA-based XL-RIS, UE will be located within the Rayleigh distance $R_{\rm ULA}=\frac{2D^2}{\lambda}$, where $D$ is the array aperture and $\lambda$ is the wavelength \cite{Rayleigh}. 
As illustrated in Fig. \ref{fig_model}, this results in the system's channel modeling exhibiting hybrid characteristics: RIS-BS channel operates in the far field, while the UE-RIS channel operates in the near field.

1) Far-field channel modeling for RIS-BS link

Under far-field communication conditions, the size of the ULA is usually much smaller than the distance between the BS and the RIS. 
In this case, the curvature difference of the spherical wave front at each antenna unit is negligible, so it can be approximated as a plane wave model and all antenna units share the same AoA $\theta$. 
Under this assumption, the wave distance difference between the RIS to the BS reference antenna (indexed as 0) and the $n$-th antenna can be expressed as
\vspace{-3mm}
\begin{equation}
    \Delta d={{r}^{(0)}}-{{r}^{(n)}}=nd\cos \theta,  n=0,1,\cdots ,N-1,
    \label{eq:3}
\end{equation}
where the antenna spacing is $d=\frac{\lambda}{2}$.
The normalized far-field array response vector ${{a}_{\chi }}(\theta )$ can be defined as
\begin{equation}
    \begin{split}
        {\mathbf{a}}_{\chi }(\theta ) & = \frac{1}{\sqrt{\chi}}[1, e^{j\frac{2\pi }{\lambda }d\cos \theta }, \cdots , e^{j\frac{2\pi }{\lambda }(\chi-1)d\cos \theta }]^{T} \\
         & = \frac{1}{\sqrt{\chi}}[1, e^{j\pi \cos \theta }, \cdots , e^{j\pi (\chi-1)\cos \theta }]^{T},
         \label{eq:4}
    \end{split}
\end{equation}
where $\chi$ denotes the number of antenna elements. 
Based on this, the channel matrix $\mathbf{G}_{\mathrm{RB}}$ between BS and RIS can be modelled as 
\begin{equation}
\mathbf{G}_{\mathrm{RB}} = \sqrt{\frac{M  N}{L_1}} \sum_{l_1=1}^{L_1} \omega_{l_1} \mathbf{a}_M\left(\varphi_{l_1}^{\text{r}_{\mathrm{RB}}}\right) \mathbf{a}_N^H\left(\phi_{l_1}^{\text{t}_{\mathrm{RB}}}\right),
\label{eq:5} 
\end{equation}
where ${{L}_{1}}$ is the number of scatterer paths between the BS and the RIS, and $\omega_{l_1}$, $\varphi _{{{l}_{1}}}^{{{r}_\mathrm{RB}}}$ and $\phi _{{{l}_{1}}}^{{{t}_\mathrm{RB}}}$ are the complex gain of the $l_1$th path, the AoA at the BS and AoD at the RIS, respectively.

2) Near-field channel modeling for UE-RIS link

In the near-field region, the planar wavefront assumption is no longer valid and an accurate modeling approach based on spherical wavefronts is required. 
In this case, the wave distance difference between the UE and the reference element of the RIS (indexed as 0) and the $n$-th element can be expressed as
\begin{equation}
    \begin{aligned}
        \Delta d'(n) &= r^{(0)} - r^{(n)} \\
                 &= r^{(0)} - \sqrt{(r^{(0)})^2 + (nd)^2 - 2 r^{(0)} nd \cos \theta} \\
                 &\approx r^{(0)} - \left( r^{(0)} - nd \cos \theta + \frac{n^2 d^2 \sin^2 \theta}{2r} \right) \\
                 &= \underbrace{nd \cos \theta}_{\text{far-field term}} - \underbrace{\frac{n^2 d^2 \sin^2 \theta}{2r}}_{\text{near-field term}}, \quad n=0,1,\cdots,N-1.
    \end{aligned}
    \label{eq:6} 
\end{equation}
On this basis, the normalized near-field array response vector $\mathbf{b}({{\theta }_{l}},{{r}_{l}})$ is a function of ${{\theta }_{l}}$ and ${{r}_{l}}$ , given by
\begin{equation}
\mathbf{b}_{\chi}({{\theta }_{l}},{{r}_{l}})=\frac{1}{\sqrt{\chi}}{{\left[ 1,{{e}^{j\frac{2\pi }{\lambda }(r_{l}^{(0)}-r_{l}^{(1)})}},\cdots ,{{e}^{j\frac{2\pi }{\lambda }(r_{l}^{(0)}-r_{l}^{(\chi-1)})}} \right]}^{T}}.
\label{eq:7}
\end{equation}
Then, the channel between the UE and the RIS is denoted as
\begin{equation}
{{\mathbf{h}}_\mathrm{UR}}=\sqrt{\frac{N}{{{L}_{2}}}}\sum\limits_{{{l}_{2}}=1}^{{{L}_{2}}}{{{\omega }_{{{l}_{2}}}}}{{\mathbf{b}}_{N}}(\phi _{{{l}_{2}}}^{{{r}_\mathrm{UR}}},r_{{{l}_{2}}}^{{{r}_\mathrm{UR}}}),
\label{eq:8}
\end{equation}
where ${{L}_{2}}$ is the number of scatterer paths between the UE and the RIS, and $\omega_{l_2}$, $\phi _{{{l}_{2}}}^{{{r}_\mathrm{UR}}}$ and $r_{{{l}_{2}}}^{{{r}_\mathrm{UR}}}$ are the complex gain of the $l_2$-th path, the AoA at the RIS, and the distance from the user to the RIS reference array element, respectively.

 3) Hybrid-field cascade channel modeling 

Since the BS and UE are respectively located in far-field  and near-field, it needs to consider the hybrid-field cascade channel characteristics under both the plane-wave and spherical-wave assumptions. 
Substituting \eqref{eq:5} and \eqref{eq:8} into \eqref{eq:1}, the cascade channel can be expressed as 
\begin{align}
\mathbf{G} 
&\triangleq 
\sqrt{\frac{M N}{L_1}} 
\sqrt{\frac{N}{L_2}} 
\sum_{l_1 = 1}^{L_1} \sum_{l_2 = 1}^{L_2} 
\omega_{l_1} \omega_{l_2} 
\mathbf{a}_M( \varphi_{l_1}^{r_\mathrm{RB}})
\nonumber \\
&\quad\cdot
\Big( 
    \mathbf{a}_N( \phi_{l_1}^{t_\mathrm{RB}}) 
    \odot 
    \mathbf{b}_N^* ( \phi_{l_2}^{r_\mathrm{UR}}, r_{l_2}^{r_\mathrm{UR}}) 
\Big)^{H}.
\label{eq:9}
\end{align}
To facilitate subsequent analysis and estimation, \textbf{Lemma 1} is introduced to simplify the model complexity as follows.

\textbf{Lemma 1:} \textit{
Let \(\mathbf{a}_N(\phi_{l_1}^{t_{\rm RB}})\) denote the far-field steering vector and
$\bm b_N(\phi_{l_2}^{r_{\rm UR}},r_{l_2}^{r_{\rm UR}})$ denote the near-field steering vector. 
Then, their Hadamard product can be approximated as}
\begin{equation}
\begin{aligned}
&\mathbf{a}_N(\phi_{l_1}^{t_\mathrm{RB}})
\odot 
\bm b_N^*(\phi_{l_2}^{r_\mathrm{UR}},r_{l_2}^{r_\mathrm{UR}})\\
&\approx \frac{1}N
\Big[\,1,\; e^{j\frac{2\pi}{\lambda}\Delta d''(1)},\;\ldots,\;
e^{j\frac{2\pi}{\lambda}\Delta d''(N-1)}\,\Big]^T,
\label{eq:10}
\end{aligned} 
\end{equation}
\textit{where the effective path difference at the \(n\)-th antenna element is given by}
\begin{equation}
\Delta {{d}^{''}}(n)=\underbrace{nd(\cos \phi _{{{l}_{1}}}^{{{t}_\mathrm{RB}}}-\cos \phi _{{{l}_{2}}}^{{{r}_\mathrm{UR}}})}_{\text{far-field term}}+\underbrace{\frac{{{n}^{2}}{{d}^{2}}{{\sin }^{2}}\phi _{{{l}_{2}}}^{{{r}_\mathrm{UR}}}}{2r_{{{l}_{2}}}^{{{r}_\mathrm{UR}}}}}_{\text{near-field term}}.
\label{eq:11}
\end{equation}
\textit{
Therefore, the Hadamard product between the far-field steering vector and the conjugated near-field steering vector can be equivalently represented by a near-field steering vector with an effective angle–distance pair, similar to \eqref{eq:6}.
}

\textit{Proof:} See Appendix A. \hfill $\blacksquare$

Based on the phase-decoupling property of \textbf{Lemma 1}, we can simplify (9) as follows:
\begin{equation}
\mathbf{G}=\sum\limits_{{{l}_{1}}=1}^{{{L}_{1}}}{\sum\limits_{{{l}_{2}}=1}^{{{L}_{2}}}{{{\beta }_{{{l}_{1}}{{l}_{2}}}}}}{{\mathbf{a}}_{M}}(\varphi _{{{l}_{1}}}^{{{r}_\mathrm{RB}}})\mathbf{b}^H_N{({{\phi }_{{{l}_{1}}{{l}_{2}}}},r_{l_2}^{r_\mathrm{UR}})},
\label{eq:12}
\end{equation}
where ${{\beta }_{{{l}_{1}},{{l}_{2}}}}\triangleq \sqrt{\frac{MN}{{L_{1}}}}\sqrt{\frac{N}{{{L}_{2}}}}{{\omega }_{{{l}_{1}}}}{{\omega }_{{{l}_{2}}}}$. 
$({{\phi }_{{{l}_{1}}{{l}_{2}}}},r_{l_2}^{r_\mathrm{UR}})$ represents the corresponding effective angle-distance pair induced by the superposition of the far-field linear phase and the near-field quadratic phase.
\subsection{Sparse Channel Estimation Problem Formulation}
Within the CS framework, the sparse representation dictionary for the original channel needs to be constructed carefully. 
In the virtual angle domain (VAD) representation, the continuous angular parameters $\varphi \in (0,\pi/2]$ and $\phi \in (-\pi,\pi)$ are discretized into a finite grid. 
To minimize the mutual coherence of the dictionary, we employ uniform sampling in the spatial frequency domain \cite{VAD uniform}.
The discrete sets of AoAs at the BS and AoDs at the RIS are respectively defined as 
\begin{align}
\Theta^{(\mathrm{BS})}
&\triangleq
\bigl\{\varphi^{(m)}:\cos\varphi^{(m)}=(m-1)/M_G\bigr\}_{m=1}^{M_G},
\label{eq:13}
\\
\Theta^{(\mathrm{RIS})}
&\triangleq
\bigl\{\phi^{(n)}:\cos\phi^{(n)}=1-2(n-1)/N_G\bigr\}_{n=1}^{N_G},
\label{eq:14}
\end{align}
where $M_G$ and $N_G$ denote the grid resolutions. Consequently, the discrete Fourier transform (DFT) array response matrix corresponding to these uniformly discretized grids are given by $\mathbf{U}_{\mathrm{BS}} = [\mathbf{a}_M(\varphi^{(1)}), \dots, \mathbf{a}_M(\varphi^{(M_G)})] \in \mathbb{C}^{M \times M_G}$ and $\mathbf{U}_{\mathrm{RIS}} = [\mathbf{a}_N(\phi^{(1)}), \dots, \mathbf{a}_N(\phi^{(N_G)})] \in \mathbb{C}^{N \times N_G}$.
We formulate the VAD representation of the cascaded channel $\mathbf{\widetilde{G}}\in \mathbb{C}^{M_G \times N}$ as follows
\begin{equation}
    \begin{aligned}
        \widetilde{\mathbf{G}} &= \mathbf{U}_{\mathrm{BS}}^{H} \mathbf{G} \mathbf{U}_{\mathrm{RIS}} \\
        &= \sum_{l_1 = 1}^{L_1} \sum_{l_2 = 1}^{L_2} \beta_{l_1 l_2} \cdot \mathbf{U}_{\mathrm{BS}}^{H} \mathbf{a}_{M} ( \varphi_{l_1}^{r_{\mathrm{RB}}} ) \mathbf{b}_{N}^{H} ( {{\phi }_{{{l}_{1}}{{l}_{2}}}},r_{l_2}^{r_{UR}} ) \mathbf{U}_{\mathrm{RIS}},
        \label{eq:15}
    \end{aligned}
\end{equation}
However, due to the spherical wavefront characteristic of near-field channel, it results in angular spreading by using $\mathbf{U}_{\mathrm{RIS}}$, which further destroys the sparse representation of $G$. 
To address this, \cite{polar} demonstrated that the near-field channels exhibit sparsity in the virtual polar domain (VPD). 
Accordingly, the VPD dictionary is constructed as
\begin{equation}
    \begin{aligned}
        \mathbf{D}_{\mathrm{RIS}} = \big[ &
            \mathbf{b}_{N}(\theta_1, r_1), \cdots, \mathbf{b}_{N}(\theta_1, r_S), \\
& \cdots, \mathbf{b}_{N}(\theta_{N_{G}}, r_1), \cdots, \mathbf{b}_{N}(\theta_{N_{G}}, r_S)
        \big] \in \mathbb{C}^{N \times N_GS},
    \end{aligned}
    \label{eq:16}
\end{equation}
where $\mathbf{b}_{N}(\theta_n, r_s)$ represents the near-field steering vector focused at the grid point $(\theta_n, r_s)$, with $\{\theta_n\}_{n=1}^{N_G}$ and $\{r_s\}_{s=1}^S$ denoting the discretized angular and distance sampling sets, respectively\footnote{The angular-domain dictionary \(\mathbf U_{\mathrm{RIS}}\) is used in DK-ODC algorithm, with \(N_G=N\) to balance complexity and estimation reliability. 
The polar-domain dictionary \(\mathbf D_{\mathrm{RIS}}\) is used in SI-VBL algorithm, with a finer angular resolution \(N_G=2N\) to enhance sparse recovery accuracy.}.
The cascaded channel is reformulated as
\begin{equation}
\mathbf{G} = \mathbf{U}_{\mathrm{BS}} \bar{\mathbf{G}} \mathbf{D}_{\mathrm{RIS}}^H,
\label{eq:17}
\end{equation}
where $\bar{\mathbf{G}}$ now represents the sparse channel matrix in the angular-polar domain. 
Substituting \eqref{eq:17} into the received signal model \eqref{eq:2}, the system model becomes
\begin{equation}
    \mathbf{Y} = \mathbf{U}_{\mathrm{BS}} \bar{\mathbf{G}} \mathbf{D}_{\mathrm{RIS}}^{H} \bm{\Gamma} + \mathbf{E}.
    \label{eq:18}
\end{equation}
Leveraging the vectorization identity $\mathrm{vec}(\mathbf{ABC}) = (\mathbf{C}^T \otimes \mathbf{A})\mathrm{vec}(\mathbf{B})$, and defining the vectorized effective received signal as $\mathbf{y} \triangleq \mathrm{vec}(\mathbf{Y}^T)$, \eqref{eq:18} can be transformed into
\begin{align}
\mathbf{y} &= \left[ \mathbf{U}_{\mathrm{BS}} \otimes (\bm{\Gamma}^T \mathbf{D}_{\mathrm{RIS}}) \right] \mathrm{vec}(\bar{\mathbf{G}}^T) + \mathrm{vec}(\mathbf{E}^T) \notag \\
&= \bm{\Phi} \mathbf{g} + \mathbf{e},
\label{eq:19}
\end{align}
where $\bm{\Phi} \in \mathbb{C}^{MT \times M_G N_GS}$ denotes the sparse representation dictionary, $\mathbf{e} \triangleq \mathrm{vec}(\mathbf{E}^T) \in \mathbb{C}^{MT \times 1}$ is the noise vector, and $\mathbf{g} \triangleq \mathrm{vec}(\bar{\mathbf{G}}^T) \in \mathbb{C}^{M_G N_GS \times 1}$ is the sparse channel vector to be estimated.

The storage overhead of $\bm{\Phi}$ increases with the product of the BS-side angular grid size and the RIS-side polar-domain grid size.
Under the considered ULA configuration, this full-dimensional dictionary requires more than 16 GB of storage, as further verified in Section~\ref{subsec:storage}.
This makes direct sparse recovery over the original cascaded dictionary impractical for memory-limited online processing platforms.
Therefore, it is essential to formulate the strategy for dimensionality reduction of sparse representation dictionary and to design an efficient sparse channel recovery algorithm.
Importantly, we employ the following hybrid-field space-time properties as guidance.

\textbf{Property 1:} \textit{In the UE-RIS near-field channel, the sparse dictionary representation of VAD maintains its group sparse property, even under the angular spreading effect.}
\begin{figure}[!t]
\centering
\includegraphics[width=2.8in]{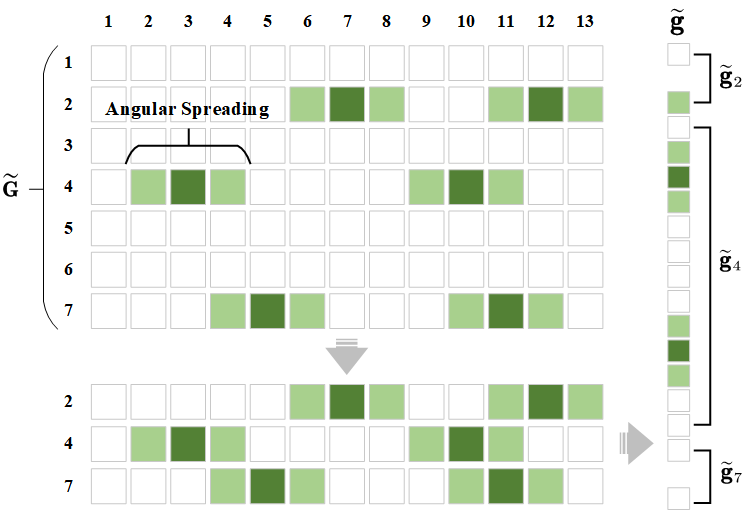}
\caption{The sparsity of cascaded channel.}
\label{Cascaded channel sparsity}
\end{figure}

As shown in Fig. \ref{Cascaded channel sparsity}, we find that only a few groups $\left\{ 2,4,7 \right\} $ determine the dominant energy of the channel. 
Subsequently, the activated groups also exhibit sparsity due to weak scatterer characteristics. 
\textbf{Property 1} indicates that the $\bm{\Phi} \in \mathbb{C}^{MT \times M_G N_GS}$ established in \eqref{eq:19} allows for sufficient compression.

\textbf{Property 2:} \textit{The cascaded channel exhibits the double-timescale characteristic. 
Since the BS, RIS, and the scatterers between them are all fixed at elevated positions, the BS–RIS channel remains quasi-static, whereas the RIS–UE channel varies rapidly due to user mobility \cite{Two-Timescale}.}

Based on the above space-time properties, it is demonstrated that under sparse representation, the VAD dictionary corresponding to the BS–RIS channel can be pre-computed and stored offline.
In contrast, the RIS–UE channel undergoes rapid fluctuations driven by user mobility, necessitating real-time online estimation of channel state. 
This double-timescale strategy effectively decouples static and dynamic components, thereby reducing storage overhead and computational complexity.


\section{Sparse Channel Representation and Recovery for Hybrid-Field}
\label{III}
In this section, we propose a phased channel estimation method for XL-RIS-assisted MIMO systems, where we have designed both sparse channel representation and recovery algorithms.
\subsection{Low-Dimensional Sparse Channel Representation for Quasi-Static RIS-BS Link}
To reduce the prohibitive dimension of the hybrid-field cascaded dictionary caused by the Kronecker product, we first develop the DK-ODC algorithm, whose main procedure is summarized in \textbf{Algorithm~\ref{alg:DK-ODC}}. 
The key idea is to exploit the quasi-static RIS-BS link to perform BS-side support detection and off-grid angular refinement only once in the offline stage. 
After this step, the full BS-side dictionary is replaced by a refined low-dimensional basis, which substantially reduces the dimension of the subsequent online sparse recovery problem.  

Specifically, by substituting \eqref{eq:15} into \eqref{eq:2} and applying the vectorization operator via the Kronecker product identity, the vectorized observational model is as follows
\begin{equation}
    \mathbf{y} = \left[ \mathbf{U}_{\mathrm{BS}} \otimes (\mathbf{\Gamma}^{T}\mathbf{U}_{\mathrm{RIS}}) \right] \mathrm{vec}(\tilde{\mathbf{G}}^T) + \mathbf{e} = \mathbf{\tilde{\Phi }}\mathbf{\tilde{g}} + \mathbf{e}.
    \label{eq:20}
\end{equation}
Through this transformation, the two-dimensional joint sparsity of the cascaded channel is mapped into a structured group sparsity within the vector space. 
Formally, we partition the vector $\tilde{\mathbf{g}}$ into $M_G$ disjoint groups, denoted as $\tilde{\mathbf{g}} = \left[ \tilde{\mathbf{g}}_1^{T}, \tilde{\mathbf{g}}_2^{T}, \ldots, \tilde{\mathbf{g}}_{M_G}^{T} \right]^{T}$, where $\tilde{\mathbf{g}}_m \in \mathbb{C}^{N_G \times 1}$ represents the $m$-th group. 
The matrix $\tilde{\mathbf{\Phi}}$ is correspondingly defined as $\tilde{\mathbf{\Phi}} = \left[ \tilde{\mathbf{\Phi}}_1, \tilde{\mathbf{\Phi}}_2, \ldots, \tilde{\mathbf{\Phi}}_{M_G} \right] \in \mathbb{C}^{MT \times M_G N_G}$, with its $m$-th sub-block given by $ \tilde{\mathbf{\Phi}}_m = \mathbf{a}_M \big( \varphi^{(m)} \big) \otimes \big( \mathbf{\Gamma}^\mathsf{T} \mathbf{U}_\mathrm{RIS} \big) \in \mathbb{C}^{MT \times N_G}$.
Physically, each resolvable path reaching the BS activates a specific group from $\mathbf{\tilde{g}}$, thereby forming a structure where only $K$ groups dominate the channel power.
The structured group sparsity of $\mathbf{\tilde{g}}$ is quantitatively characterized by the mixed $\ell_{1,0}$ pseudo-norm, defined as
\begin{equation}
    \|\mathbf{\tilde{g}}\|_{1,0} = \sum_{m=1}^{M_G} \mathbb{I}\left( \|\mathbf{\tilde{g}}_{m}\|_1 > \epsilon \right) = K \ll M_G,
    \label{eq:21}
\end{equation}
where $\epsilon$ denotes a noise-dependent threshold.
Based on the group sparsity property, the group support set can be obtained using an orthogonal projection strategy. 
As for $\mathbf{\tilde{g}}$, its non-zero group support can be updated as 
\begin{equation}
    {{\mathcal{M}}}^{(k)}={{\mathcal{M}}^{(k-1)}}\cup \{{{\hat{m}}^{(k)}}\},|\mathcal{M}|=K\ll {{M}_{G}},
    \label{eq:22}
\end{equation}
That is, eventually $|\mathcal{M}|$ contains all the non-zero group indices about $\mathbf{\tilde{g}}$. 
${{\hat{m}}^{(k)}}$ is the maximum correlation index, the index with the highest correlation is most likely to be the non-zero group in the $k$-th iteration, which is given by
\begin{equation}
\hat{m}^{(k)}
=
\underset{m \in \{1,\ldots,M_G\}}{\arg\max}
\;
\left\|
\mathbf{\tilde{\Phi}}_{m}^{H}
\mathbf{\widetilde{r}}_{[\mathcal{M}]}^{(k-1)}
\right\|_{1},
\label{eq:23}
\end{equation}
where $\mathbf{\tilde{r}}_{{\mathcal{M}}}^{(k-1)}=\mathbf{y}-\mathbf{\tilde{\Phi }}[:,{{\mathcal{M}}^{(k-1)}}]\mathbf{\tilde{g}}_{{[\mathcal{M}]}}^{(k-1)}\in {{\mathbb{C}}^{MT\times 1}}$ is the residual vector obtained from the previous iteration. 
The channel coefficient $\mathbf{\tilde{g}}_{[\mathcal{M}]}^{(k-1)}$ is determined by the least squares (LS) solution
\begin{equation}
\mathbf{\tilde{g}}_{[\mathcal{M}]}^{(k-1)} = \left( \mathbf{\tilde{\Phi}}_{[\mathcal{M}]^{(k-1)}}^H \mathbf{\tilde{\Phi}}_{[\mathcal{M}]^{(k-1)}} \right)^{-1} \mathbf{\tilde{\Phi}}_{[\mathcal{M}]^{(k-1)}}^H \mathbf{y},
\label{eq:24}
\end{equation}
where $\mathbf{\tilde{\Phi}}_{\mathcal{M}^{(k-1)}} \triangleq \mathbf{\tilde{\Phi} }[:,{\mathcal{M}}^{(k-1)}]\in {{\mathbb{C}}^{MT\times |{{\mathcal{M}}^{(k-1)}}|{{N}_{G}}}}$ is the submatrix of $\mathbf{\tilde{\Phi }}$ containing the support set $\mathcal{M}$.
The most relevant index of the BS side uniformly discrete set ${{\Theta }^{(\mathrm{BS})}}$ with respect to the true AoAs is given by \eqref{eq:23}, which provides a rough estimate of the grid points for each AoA, i.e., $\Theta _{{\hat{m}}}^{(n)}=(\hat{m}-1)/{{M}_{G}},\text{ }\hat{m}\in 1,2,\cdots ,{{M}_{G}}$. 
Since the true AoAs grid values are located around the roughly divided grid and depend on the nearest grid point in the dictionary. 
However, the conventional on-grid representation inherently suffers from off-grid error caused by the inevitable mismatch between the continuous true AoAs and the discrete grid points \cite{Cao_Efficient}. 
To quantify this mismatch and facilitate high-precision estimation, we first characterize the energy leakage properties of the array response in the following proposition.

 \textbf{Proposition 1:} 
 \textit{Based on the definition in \eqref{eq:15}, let $\mathbf{z}=\mathbf{U}_{\mathrm{BS}}^{\mathrm{H}}{{\mathbf{a}}_{M}}(\varphi _{{{l}_{1}}} ^{{{r}_{\mathrm{RB}}}})$ denote the projection of the array response vector onto the DFT dictionary. 
 The $\hat{m}$-th entry of the vector $\mathbf{z}\in {{\mathbb{C}}^{{{M}_{G}}\times 1}}$ is derived as}
\begin{align}
    {{z}_{\hat{m}}} &= {{\mathbf{a}}_{M}}{{({{\varphi }^{\hat{m}}})}^{\mathrm{H}}}{{\mathbf{a}}_{M}}(\varphi _{{{l}_{1}}}^{{{r}_{\mathrm{RB}}}}) \notag \\
    &= \frac{1}{M} {{e}^{-j\frac{\pi (M-1)}{\lambda }d{{\Delta }_{m}}}}\cdot {{D}_{M}}({{\Delta }_{m}}),
    \label{eq:25}
\end{align}
\textit{where ${{\Delta }_{{m}}}=\cos ({{\varphi }^{\hat{m}}})-\cos (\varphi _{{{l}_{1}}}^{{{r}_{\mathrm{RB}}}})$ represents the directional cosine difference, and ${{D}_{M}}(\cdot)$ is the Dirichlet kernel defined as \cite{Dirichlet_kernel}}
\begin{equation}
    {{D}_{M}}({{\Delta }_{m}})=\frac{\sin \left( \frac{\pi M d}{\lambda} {{\Delta }_{m}} \right)}{\sin \left( \frac{\pi d}{\lambda} {{\Delta }_{m}} \right)}.
    \label{eq:26}
\end{equation}
\textit{If the angle $\varphi _{{{l}_{1}}}^{{{r}_{RB}}}$ coincides exactly with a grid point ${{\varphi }^{(\hat{m})}}$, the magnitude peaks at ${{z}_{{{m}'}}}=1$. 
Otherwise, $|z_m|$ exhibits sidelobe attenuation governed by the Dirichlet kernel envelope, which decays approximately as
$1/|\Delta_m|$ away from the main lobe.}

\textit{Proof:} See Appendix B. \hfill $\blacksquare$

\noindent\textit{Remark 1 (Closely spaced BS-side paths):}
In the multi-path case, the projected spectrum is a superposition of shifted Dirichlet kernels.
Hence, closely spaced paths are inherently difficult to resolve because their main lobes and sidelobes may overlap and shift the observed local peaks. 
This is a fundamental finite-aperture effect in array processing and off-grid line spectral estimation \cite{ANM}.

\begin{figure}[!t]
\centering
\includegraphics[width=3in]{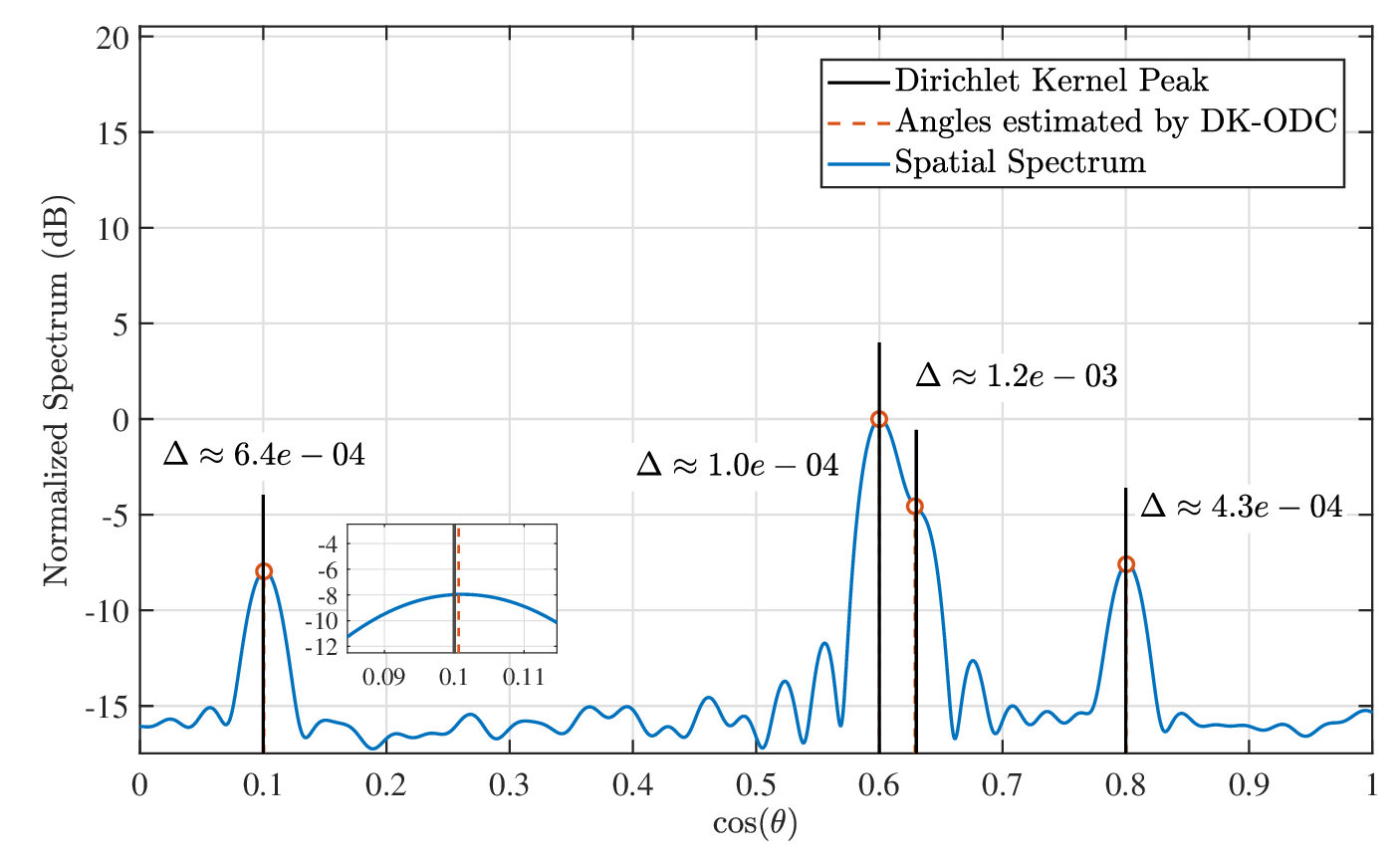}
\caption{Off-grid estimation performance, SNR = 0 dB, $T=80, M=64, N=256$.}
\label{angles estimation}
\end{figure}

Let \(u_{l_1}=\cos\varphi_{l_1}^{\mathrm r_{\mathrm{RB}}}\) denote the
BS-side spatial frequency, and define
\begin{equation}
    \Delta_{\rm BS}
    \triangleq
    \min_{l_1\neq l_1'}
    \left|u_{l_1}-u_{l_1'}\right|.
\end{equation}
When \(\Delta_{\rm BS}\) is smaller than approximately \(2/M\), the kernel overlap becomes more severe, and two close paths may be merged into one effective angular support. 
Fig.~\ref{angles estimation} therefore includes a representative close-spacing realization to examine whether DK-ODC can still detect resolvable dominant paths under Dirichlet-kernel overlap.

Motivated by \textbf{Proposition 1}, the true continuous angle estimation translates into Dirichlet kernel peak estimation.
Given the estimated grid point $\Theta _{\hat{m}}^{(k)}$, the search region for the peak ${{\hat{\Theta }}^{(n)}}$ is restricted to the interval $[\Theta _{d}^{(k)}, \Theta _{u}^{(k)}]$, rather than the entire angular domain. 
The boundaries are determined by the midpoints of adjacent grid points: $\Theta _{d}^{(k)}=[\Theta _{{\hat{m}}}^{(k)}+\Theta _{\hat{m}-1}^{(k)}]/2$ and $\Theta _{u}^{(k)}=[\Theta _{\hat{m}+1}^{(k)}+\Theta _{{\hat{m}}}^{(k)}]/2$, where $\Theta _{\hat{m}\pm 1}^{(k)}$ represent the neighbors of $\hat{m}$. 
Accordingly, the refined support set $\Omega^{(k)}$ is formulated as
\begin{equation}
\Omega^{(k)}
\triangleq
\{
\tilde{\Theta}^{(i)}
\}_{i=1}^{k}
=
\Omega^{(k-1)}
\cup
\{
\tilde{\Theta}^{(k)}
\}.
\label{eq:27}
\end{equation}
Consequently, the full dictionary $\mathbf{U}_{\mathrm{BS}}$ is downsized to construct a reduced dictionary $\mathbf{\hat{U}}_{\mathrm{BS}} = [\mathbf{a}(\Omega_1), \dots, \mathbf{a}(\Omega_K)] \in \mathbb{C}^{M \times K}$, which retains only the array response vectors corresponding to the refined support set $\Omega^{(k)}$.
Based on this reduced basis, the fine-grained angle estimation for each scattering path is formulated as a local optimization problem searching for the Dirichlet kernel peak $\tilde{\Theta }$
\begin{equation}
\begin{aligned}
\tilde{\Theta}^{(k)}
= \;& \underset{\tilde{\Theta} \in \mathcal{I}^{(k)}}{\arg \min}
\; \mathcal{J}\!\left(\tilde{\Theta}\right)
\;\triangleq\;
\left\|
\mathbf{y}
-
\widetilde{\mathbf{\Phi}}_{[\mathcal{M}]}^{(k)}\!\left(\tilde{\Theta}\right)
\mathbf{\tilde{g}}_{[\mathcal{M}]}^{(k)}
\right\|_{2}^{2},
\\
& \mathcal{I}^{(k)} \triangleq
\left[ \Theta_{d}^{(k)}, \, \Theta_{u}^{(k)} \right].
\label{eq:28}
\end{aligned}
\end{equation}

This problem is solved using the alternating minimization (AM) algorithm over the bounded interval, as detailed in Appendix  C.  \hfill $\blacksquare$

As shown in Fig.~\ref{angles estimation}, the considered realization contains four BS-side spatial frequencies, \(u=\{0.1,0.6,0.63,0.8\}\), where \(0.6\) and \(0.63\) form a closely spaced pair. 
The estimated peaks marked by red circles match the true ones with errors on the order of \(10^{-3}\), verifying the off-grid refinement capability of DK-ODC algorithm under resolvable close-spacing conditions.
Crucially, the construction of the truncated dictionary $\mathbf{\hat{U}}_{\mathrm{BS}}$ functions as an offline dimensionality reduction strategy, compressing the problem size prior to online iterations. 
Facilitated by this reduction and the inherent spatial sparsity
of the UE--RIS near-field channel, the compressed sensing and
reconstruction dictionaries are respectively given by
\begin{equation}
\tilde{\boldsymbol{\Phi}}^{\downarrow}_{P}
=
\hat{\mathbf U}_{\mathrm{BS}}
\otimes
\left(
\boldsymbol{\Gamma}^{\mathrm T}
\mathbf D_{\mathrm{RIS}}
\right)
\in
\mathbb C^{MT \times K N_G S},
\label{eq:29}
\end{equation}
\begin{equation}
\tilde{\boldsymbol{\Psi}}^{\downarrow}_{P}
=
\hat{\mathbf U}_{\mathrm{BS}}
\otimes
\mathbf D_{\mathrm{RIS}}
\in
\mathbb C^{MN \times K N_G S}.
\label{eq:rec_dict}
\end{equation}
Here, $\tilde{\boldsymbol{\Phi}}^{\downarrow}_{P}$ is used for pilot-domain Bayesian inference, while
$\tilde{\boldsymbol{\Psi}}^{\downarrow}_{P}$ is used for
cascaded-channel reconstruction.
The dictionary dimension is reduced from $M_GN_GS$ to $KN_GS$, $K \ll M_G$.
Thus, to estimate the entire cascade channel, the system model is reformulated as
\begin{equation}\mathbf{y} = \mathbf{\tilde{\Phi}}^{\downarrow}_{P} \mathbf{\tilde{g}}^{\downarrow} + \mathbf{e},
\end{equation}
where $\mathbf{\tilde{g}}^{\downarrow} = [\hat{g}_1, \ldots, \hat{g}_{KN_GS}]^T$ denotes the channel coefficients to be estimated. 
DK-ODC algorithm can also be computed offline to support the subsequent online Bayesian recovery process described in Section \ref{subsec:SI-VBL}.
    
    
\begin{algorithm}[!t]
    \caption{The Proposed DK-ODC Algorithm}
    \label{alg:DK-ODC}
    \begin{algorithmic}[1]

        \Require Observation $\mathbf y$, initial sensing dictionary
        $\tilde{\boldsymbol{\Phi}}$, polar dictionary
        $\mathbf D_{\rm RIS}$, training matrix $\boldsymbol{\Gamma}$,
        tolerance $\epsilon$, $K$.

        \State \textbf{Initialize:}
        $k=1$, $\mathbf r^{(0)}=\mathbf y$,
        $\mathcal M^{(0)}=\varnothing$,
        $\boldsymbol{\Omega}^{(0)}=\varnothing$.
        
        \While{$k\le K$ and $\|\mathbf r^{(k-1)}\|_2>\epsilon$}
            \State Detect the coarse BS-side group index
            $\hat m^{(k)}$ by \eqref{eq:23}.
            
            \State Update
            $\mathcal M^{(k)}
            \leftarrow
            \mathcal M^{(k-1)}\cup\{\hat m^{(k)}\}$.
            
            \State Refine $\tilde{\Theta}^{(k)}$ by solving
            \eqref{eq:28}, and update
            $\boldsymbol{\Omega}^{(k)}
            \leftarrow
            \boldsymbol{\Omega}^{(k-1)}\cup\{\tilde{\Theta}^{(k)}\}$.
            
            \State Estimate $\tilde{\mathbf g}^{(k)}_{\mathcal M}$ by
            \eqref{eq:24}, and update
            $\mathbf r^{(k)}
            =
            \mathbf y
            -
            \tilde{\boldsymbol{\Phi}}_{\mathcal M^{(k)}}
            \tilde{\mathbf g}^{(k)}_{\mathcal M}$.
            
            \State $k\leftarrow k+1$.
        \EndWhile
        
        \State Construct
        $\hat{\mathbf U}_{\rm BS}
        =
        [\mathbf a_M(\Omega_1),\ldots,
        \mathbf a_M(\Omega_{K})]$.
        
        \State Form
        $\tilde{\boldsymbol{\Phi}}_{P}^{\downarrow}
        =
        \hat{\mathbf U}_{\rm BS}
        \otimes
        (\boldsymbol{\Gamma}^{T}\mathbf D_{\rm RIS})$
        and
        $\tilde{\boldsymbol{\Psi}}_{P}^{\downarrow}
        =
        \hat{\mathbf U}_{\rm BS}
        \otimes
        \mathbf D_{\rm RIS}$.
        
        \State \Return
        $\tilde{\boldsymbol{\Phi}}_{P}^{\downarrow}$,
        and
        $\tilde{\boldsymbol{\Psi}}_{P}^{\downarrow}$.
    \end{algorithmic}
\end{algorithm}
\subsection{Efficient Sparse Channel Recovery for Dynamic UE-RIS link}
\label{subsec:SI-VBL}
Although DK-ODC reduces the dictionary dimension from $M_GN_GS$ to $KN_GS$ by removing redundant BS-side angular components, directly applying conventional VBL to the compressed dictionary $\mathbf{\tilde{\Phi}}^{\downarrow}_{P}\in\mathbb{C}^{MT\times KN_GS}$ remains computationally demanding. 
This is because each VBL iteration requires inverting a posterior covariance matrix of size $(KN_GS)\times(KN_GS)$, whose complexity is  $\mathcal{O}\!\left((KN_GS)^3\right)$.
To address this issue, we develop the SI-VBL algorithm, whose detailed steps are summarized in \textbf{Algorithm~\ref{alg:SI-VBL}}.

Specifically, we exploit the VPD sparsity of the UE-RIS channel, where signal energy is naturally concentrated in a few basis vectors. 
Rather than performing Bayesian learning on the full dictionary \cite{SBL}, we project the observation signal onto the dictionary atoms to assess their energy contribution. 
This effectively solves the channel estimation problem in low-dimensional subspaces, while accelerating computation and suppressing noise-dominated components. 

Formally, the correlation energy vector $\mathbf{p} \in \mathbb{R}^{KN_GS \times 1}$ is obtained by projecting the observation $\mathbf{y}$ onto each column of $\mathbf{\tilde{\Phi}}^{\downarrow}$
\begin{equation}
p_i = | ( \mathbf{\tilde{\Phi}}^{\downarrow}_{:,i})^H \mathbf{y} |^2_2, \quad i = 1, \ldots, KN_GS.
\label{eq:31}
\end{equation}
To construct the dominant subspace, we select the indices corresponding to the $Q$ largest energy peaks. 
Let $\mathcal{I}_{sub}$ denote the index set of the retained atoms, defined as
\begin{equation}
    \mathcal{I}_{\mathrm{sub}} = \left\{ i \in \{1, \ldots, KN_GS\} \mid p_i \ge p_{(Q)} \right\},
    \label{eq:I_sub}
\end{equation}
where $p_{(Q)}$ represents the $Q$-th largest value in the vector $\mathbf{p}$. 
The cardinality of this set is $|\mathcal{I}_{sub}| = Q \ll KN_GS$. 
Accordingly, we construct the subspace dictionary $\mathbf{\Xi} =[\bm{\psi}(\theta_1, r_1), \ldots, \bm{\psi}(\theta_Q, r_Q)]\in \mathbb{C}^{MT \times Q}$ by extracting the columns of $\mathbf{\tilde{\Phi}}^{\downarrow}$ indexed by $\mathcal{I}_{sub}$, i.e., $\mathbf{\Xi} = \mathbf{\tilde{\Phi}}^{\downarrow}_{[:, \mathcal{I}_{sub}]}$. 
The original high-dimensional optimization problem is thus compressed into this low-dimensional subspace, and the system model is reformulated as
\begin{equation}
\mathbf{y} = \mathbf{\Xi} \mathbf{w} + \mathbf{e},
\end{equation}
where $\mathbf{w} \in \mathbb{C}^{Q \times 1}$ denotes the channel weights associated with the active subspace.
By constraining the optimization domain within $\mathbf{\Xi}$ and employing the variational Bayesian learning (VBL) framework \cite{2024_Variational}, the channel weights are dynamically optimized within the subspace, thereby achieving high-precision channel recovery.

In the VBL framework, the joint probability density function is assumed to decompose as \cite{FVBL}, \cite{2024_Super}
\begin{equation}
    p(\textbf{w},\sigma , \bm{\alpha} ,\mathbf{y})=p(\mathbf{y}|\textbf{w},\sigma )p(\textbf{w}|\bm{\alpha} )p(\bm{\alpha} )p(\sigma ).
\end{equation}
Under the Gaussian noise assumption, the likelihood $p(\mathbf{y}|\textbf{w},\sigma )$ is given as $p(\mathbf{y}|\textbf{w},\sigma )=\mathcal{CN}(\mathbf{y}|\mathbf{\Xi} \textbf{w},{{\sigma }^{-1}}\textbf{I})$. 
The sparse prior \(p(\mathbf w|\boldsymbol{\alpha})\) is decomposed into $p(\textbf{w}|\bm{\alpha} )=\underset{q=1}{\overset{Q}{\mathop \prod }}\,p({{w}_{q}}|{{\alpha }_{q}})$ , where $p({{w}_{q}}|{{\alpha }_{q}})=\mathcal{CN}({{w}_{q}}|0,\alpha _{q}^{-1})$. 
The choice of prior $p(\sigma )$ is the gamma distribution, i.e., $p(\sigma )=Ga(\sigma |c,d)$, since it is the conjugate prior of the accuracy of the Gaussian likelihood $p(\mathbf{y}|\textbf{w},\sigma )$. 
The prior $p({{\alpha }_{q}})$ also referred to as the super-prior of the $q$-th component is chosen to be $Ga({{\alpha }_{q}}|{a_{q}},{b_{q}})$ \cite{FVBL}. 
Thus the set of unknown parameters to be estimated is defined as $\boldsymbol{\vartheta} =\left\{ \textbf{w},\bm{\alpha} ,\sigma  \right\}$.
In this case, sparsity and channel estimation are jointly considered if we can compute the maximum a posteriori (MAP) estimate of $\boldsymbol{\vartheta} $ for the given $\mathbf{y}$. 
However, accurate MAP estimation of $\boldsymbol{\vartheta} $ is difficult due to the multidimensional integral over the marginal distribution of $\mathbf{y}$ involved. 
Therefore, we use the VBL method \cite{VBI} to approximate the posterior distribution under the variational principle and Bayesian inference. 
The approach can be adapted to infer the MAP of the unknown parameters in $\boldsymbol{\vartheta} $. 
Specifically, instead of directly estimating the posterior $P(\boldsymbol{\vartheta} |\mathbf{y})$ of $\boldsymbol{\vartheta} $, $P(\boldsymbol{\vartheta} |\mathbf{y})$ is approximated using the decomposed posterior $q(\text{ }\!\!\boldsymbol{\vartheta}\!\!\text{ })$. 
The difference between two distributions $q(\text{ }\!\!\boldsymbol{\vartheta}\!\!\text{ })$ and $P(\boldsymbol{\vartheta} |\mathbf{y})$ is calculated by minimizing the corresponding Kullback-Leibler (KL) divergence, i.e.,
\begin{equation}
    {{D}_{KL}}(q(\boldsymbol{\vartheta} )||P(\boldsymbol{\vartheta} |\textbf{y)})={{\left\langle \ln \frac{q(\boldsymbol{\vartheta} )}{P(\boldsymbol{\vartheta} |\mathbf{y})} \right\rangle }_{q(\boldsymbol{\vartheta} )}},
\end{equation}
or equivalently maximize the evidence lower bound (ELBO), i.e.,
\begin{equation}
    q(\boldsymbol{\vartheta} )=\underset{q(\boldsymbol{\vartheta} )}{\mathop{\arg \max }}\,\underbrace{\int{q(\boldsymbol{\vartheta} )}\ln \frac{p(y,\boldsymbol{\vartheta} )}{q(\boldsymbol{\vartheta} )}d\boldsymbol{\vartheta} }_{\triangleq  Q(q(w)q(\alpha )q(\sigma ))}.
\end{equation}
Suppose that $q(\text{ }\!\!\boldsymbol{\vartheta}\!\!\text{ })$ is factored into
\begin{equation}
    q(\text{ }\!\!\boldsymbol{\vartheta}\!\!\text{ })\triangleq q\textbf{(w})q(\bm{\alpha} )q(\sigma ),
\end{equation}
where $q(\cdot )$ denotes the posterior factor. 
The ELBO maximization of $q(\text{ }\!\!\boldsymbol{\vartheta}\!\!\text{ })$ can be obtained by using the Lagrange multiplier method \cite{JQ}, which leads to an alternating approach, i.e.,
\begin{equation}
    \ln q({{\text{ }\!\!\boldsymbol{\vartheta}\!\!\text{ }}_{n}})\propto {{\left\langle \ln p(\mathbf{y},\text{ }\!\!\boldsymbol{\vartheta}\!\!\text{ }) \right\rangle }_{q(\text{ }\!\!\boldsymbol{\vartheta}\!\!\text{ })/q({{\text{ }\!\!\boldsymbol{\vartheta}\!\!\text{ }}_{n}})}},n=1,2,3,
    \label{eq:38}
\end{equation}
where ${{\text{ }\!\!\boldsymbol{\vartheta}\!\!\text{ }}_{n}}$ denotes the $n$-th element of the $\boldsymbol{\vartheta} $, ${{\left\langle \cdot  \right\rangle }_{q(\boldsymbol{\vartheta} )/q({{\boldsymbol{\vartheta} }_{n}})}}$ denotes the expectation of $q(\text{ }\!\!\boldsymbol{\vartheta}\!\!\text{ })/q({{\text{ }\!\!\boldsymbol{\vartheta}\!\!\text{ }}_{n}})$ for the distribution $q(\boldsymbol{\vartheta} )$ without $q({{\boldsymbol{\vartheta} }_{n}})$ terms, the following derives the details of the update of the variables in $\boldsymbol{\vartheta} $.\\

1) \underline{\textbf{Update of $q(\bm{\alpha} )$:}} By substituting the priors $p(\textbf{w}|\bm{\alpha} )$ and $p(\bm{\alpha} )$ into \eqref{eq:38}, we obtain
\begin{equation}
    \ln q(\bm{\alpha} )\propto {{\left\langle \ln p(\textbf{w}|\bm{\alpha} )+\ln p(\bm{\alpha} ) \right\rangle }_{q(\textbf{w})}}.
\end{equation}
Following the analysis in \cite{FVBL}, and given the factorization $\ln q(\bm{\alpha} )=\sum_{q=1}^{Q}\ln q({{\alpha }_{q}})$, we can determine the stationary points for the variational parameters of each factor $q({{\alpha }_{q}})$. 
This establishes a dependency between the ELBO and the specific spatial parameters (angle and distance) of a basis function. 
Specifically, for a given basis function $\bm{\psi} (\theta_q, r_q)$, the sequence $\left\{ {{{\hat{\alpha }}}_{q}} \right\}_{q=1}^{Q}$, obtained by alternating updates of $q(\textbf{w})$ and $q({{\alpha }_{q}})$, converges to the fixed point
\begin{equation}
\hat{\alpha}_{q} = 
\begin{cases}
    \left( w_{q}^{2} - \varsigma_{q} \right)^{-1}, & \text{if }  \frac{w_{q}^{2}}{\varsigma_{q}} > 1+\kappa, \\ 
    \infty, & \text{if } \frac{w_{q}^{2}}{\varsigma_{q}} \leq 1+\kappa,
    \label{eq:40}
\end{cases}
\end{equation}
where $\kappa$ denotes pruning threshold and the pruning statistics ${{\varsigma }_{q}}$ and ${{w}_{q}}$ are defined as
\begin{multline}
\varsigma_{q} = \left( \bm{\psi}(\theta_{q}, r_q)^{H} \hat{\sigma} \bm{\psi}(\theta_{q}, r_q) 
- \bm{\psi}(\theta_{q}, r_q)^{H} \hat{\sigma} \mathbf{\Xi}_{-q} \bm{\Sigma}_{-q} \right. \\
\left. \times \mathbf{\Xi}_{-q}^{H} \hat{\sigma} \bm{\psi}(\theta_{q}, r_q) \right)^{-1},
\label{eq:41}
\end{multline}
\begin{equation}
w_q^2
=
\left|
\varsigma_q
\boldsymbol{\psi}^H(\theta_q,r_q)
\hat{\sigma}
\bar{\mathbf r}_q
\right|^2,
\label{eq:42}
\end{equation}
where $\mathbf{\Xi}_{-q}$ represents the dictionary excluding the $q$-th component. 
The associated terms are detailed as follows:
\begin{align}
  & \hat{\textbf{A}}_{-q} = \text{diag}\left( \left[ \hat{\alpha}_{1}, \ldots, \hat{\alpha}_{q-1}, \hat{\alpha}_{q+1}, \ldots, \hat{\alpha}_{Q} \right] \right), \notag\\ 
  & \mathbf{\Xi}_{-q} = \left[ \bm{\psi}(\theta_{1}, r_1), \ldots, \bm{\psi}(\theta_{q-1}, r_{q-1}), \bm{\psi}(\theta_{q+1}, r_{q+1}), \ldots \right], \notag\\ 
  & \bm{\Sigma}_{-q} = \left( \mathbf{\Xi}_{-q}^{H} \hat{\sigma} \mathbf{\Xi}_{-q} + \hat{\textbf{A}}_{-q} \right)^{-1},\\ 
  & \bm{\mu}_{-q} = \bm{\Sigma}_{-q} \mathbf{\Xi}_{-q}^{H} \hat{\sigma} \,\mathbf{y}, \notag\\ 
  & \bar{\textbf{r}}_{q} = \mathbf{y} - \mathbf{\Xi}_{-q} \bm{\mu}_{-q}. \notag
\end{align}
Let $\mathcal S\subseteq \mathcal I_{\rm sub}$ denote the active support set retained after the pruning step, and let $R=|\mathcal S|$ be its cardinality. 
Due to the sparsity of the UE--RIS channel, only a small subset of candidate atoms remains active, leading to $R\ll Q\ll KN_GS$. 
The Bayesian posterior update is therefore carried out over the active dictionary
$\boldsymbol{\Xi}_{\mathcal S}
=
\boldsymbol{\Xi}[:,\mathcal S]
\in\mathbb C^{MT\times R}$.
This reduces the dimension of the covariance update from $Q$ to $R$, thereby avoiding unnecessary posterior updates over inactive atoms.

Result \eqref{eq:40} provides the criterion for basis pruning. The algorithm sequentially evaluates the $q$-th candidate pair $(\theta_q, r_q)$ to determine its validity
\begin{itemize}
    \item If $\frac{w_{q}^{2}}{\varsigma_{q}}>1+\kappa$, then $\hat{\alpha}_{q}$ is set to a finite value defined by $\hat{\alpha}_{q}=(w_{q}^{2}-{\varsigma }_{q})^{-1}$. 
    Consequently, the basis function $\bm{\psi}(\theta_q, r_q)$ is retained, and the dictionary is updated as $\mathbf{\Xi}^{(q+1)}=[\mathbf{\Xi}^{(q)},\bm{\psi} (\theta_{q}, r_q)]$ with hyperparameters $\hat{\textbf{A}}^{(q+1)}=[\hat{\textbf{A}}^{(q)},\hat{\alpha}_{q}]$.
    \item If $\frac{w_{q}^{2}}{\varsigma_{q}} \leq 1+\kappa$, then $\hat{\alpha}_{q} \to \infty$, indicating the component is redundant and thus pruned.
\end{itemize}
\noindent\textit{Remark 2:}
The pruning threshold $\kappa$ in \eqref{eq:40} is motivated by marginal log-likelihood maximization \cite{FVBL}. 
For a candidate basis $\bm{\psi}(\theta_q,r_q)$, its likelihood contribution can be separated as
\begin{equation}
\Delta\mathcal{L}(\alpha_q)
=
\frac{1}{2}
\left(
\ln\frac{\alpha_q}{\alpha_q+\varsigma_q}
+
\frac{w_q^2}{\alpha_q+\varsigma_q}
\right),
\end{equation}
where $\varsigma_q$ denotes the uncertainty cost and $w_q^2$ denotes the residual projection energy. 
Maximizing $\Delta\mathcal{L}(\alpha_q)$ yields 
$\hat{\alpha}_q=(w_q^2-\varsigma_q)^{-1}$, which is feasible only if $w_q^2/\varsigma_q>1+\kappa$. 
Hence, $\kappa$ balances false-alarm suppression and weak-path preservation.

2) \underline{\textbf{Update of $q(\textbf{w})$:}} By substituting $p(\textbf{w}|\bm{\alpha} )$ prior distributions into \eqref{eq:38}, we obtain
\begin{align}
    \ln q(\textbf{w}) &\propto \left\langle \ln p(\mathbf{y}|\textbf{w},\sigma) + \ln p(\textbf{w}|\bm{\alpha}) \right\rangle_{q(\bm{\alpha}) q(\sigma)} \notag \\
    &\propto -\frac{1}{2} \textbf{w}^{H} \bm{\Sigma}^{-1} \textbf{w} - \textbf{w}^{H} \bm{\Sigma}^{-1} \bm{\mu}.
\end{align} 
It shows that $q(\textbf{w})$ follows a Gaussian distribution, i.e., $q(\textbf{w})\sim \mathcal{C}\mathcal{N}(\textbf{w}|\bm{\mu} ,\bm{\Sigma} )$, and its covariance and mean are as follows
\begin{equation}
    \bm{\Sigma}^{(q+1)} ={{(\hat{\sigma }{\boldsymbol{\Xi}_{\mathcal S}}^{H}}\boldsymbol{\Xi}_{\mathcal S}}+\hat{\textbf{A}})^{-1},
    \label{eq:45}
\end{equation}
\begin{equation}
    \bm{\mu}^{(q+1)} =\hat{\sigma }\bm{\Sigma} {\boldsymbol{\Xi}_{\mathcal S}}^{H}\mathbf{y}.
    \label{eq:46}
\end{equation}

3) \underline{\textbf{Update of $q(\sigma)$:}} By substituting $p(\sigma)$ prior distributions into \eqref{eq:38}, we obtain
\begin{align}
 \ln q(\sigma )&\propto {{\left\langle \ln p(\mathbf{y}|\textbf{w},\sigma )+\ln p(\sigma ) \right\rangle }_{q(\textbf{w})}} \notag\\ 
 &\propto MT\log \sigma -\sigma \left( ||\mathbf{y}-\boldsymbol{\Xi}_{\mathcal S}\bm{\mu} ||_{2}^{2}+tr(\bm{\Sigma} {\boldsymbol{\Xi}_{\mathcal S}}^{H}\boldsymbol{\Xi}_{\mathcal S}) \right).
\end{align}
Therefore, $q(\sigma )$ follows a Gamma distribution in which $\hat{\sigma }$ is given by
\begin{equation}
    \hat{\sigma }=\frac{MT+c_1}{||\mathbf{y}-\boldsymbol{\Xi}_{\mathcal S}\bm{\mu} ||_{2}^{2}+tr(\bm{\Sigma}{\boldsymbol{\Xi}_{\mathcal S}}^{H}\boldsymbol{\Xi}_{\mathcal S})+d_1},
    \label{eq:48}
\end{equation}
where $c_1=d_1=1e-4$. 
In this framework, the goal is to determine whether a candidate basis $\bm{\psi}(\theta_q, r_q)$ contributes sufficiently to the active dictionary $\boldsymbol{\Xi}_{\mathcal S}$. 

In summary, the proposed framework first uses DK-ODC to exploit the quasi-static RIS-BS link and compress the sparse representation from $M_GN_GS$ to $KN_GS$. 
Then, SI-VBL performs online recovery over the compressed polar-domain dictionary by screening a $Q$-dimensional candidate subspace and pruning it to an $R$-dimensional active subspace, where $R\ll Q\ll KN_GS$. 
Thus, the Bayesian posterior update is restricted to the active dictionary instead of the full compressed dictionary. 
To provide a clear understanding of the entire algorithm, the overall channel estimation framework is summarized in \textbf{Algorithm \ref{alg:overall_framework}}.

\begin{algorithm}[!t]
    \caption{The Proposed SI-VBL Algorithm}
    \label{alg:SI-VBL}
    \begin{algorithmic}[1]
        \Require Observation $\mathbf y$, compressed sensing dictionary
        $\tilde{\boldsymbol{\Phi}}_{P}^{\downarrow}$, compressed
        reconstruction dictionary $\tilde{\boldsymbol{\Psi}}_{P}^{\downarrow}$,
        subspace size $Q$, pruning threshold $\kappa$, tolerance $\varepsilon$.

        \State \textbf{Initialize:}
        $\mathcal S=\varnothing$,
        $\boldsymbol{\Xi}_{\mathcal S}=\varnothing$,
        $\boldsymbol{\Psi}_{\mathcal S}=\varnothing$,
        $\hat{\mathbf A}=\varnothing$,
        $\hat{\sigma}=1$.

        \State Compute the correlation energy vector $\mathbf p$ by \eqref{eq:31}.
        \State Construct the candidate index set $\mathcal I_{\rm sub}$ by \eqref{eq:I_sub}.
        \State Form
        $\boldsymbol{\Xi}
        =\tilde{\boldsymbol{\Phi}}_{P}^{\downarrow}[:,\mathcal I_{\rm sub}]$
        and
        $\boldsymbol{\Psi}
        =\tilde{\boldsymbol{\Psi}}_{P}^{\downarrow}[:,\mathcal I_{\rm sub}]$.

        \For{$q=1,\ldots,Q$}
            \State Let $\boldsymbol{\phi}_q=\boldsymbol{\Xi}_{:,q}$ and
            $\boldsymbol{\psi}_q=\boldsymbol{\Psi}_{:,q}$.
            \State Compute $\varsigma_q$ and $w_q^2$ by
            \eqref{eq:41} and \eqref{eq:42}, respectively.

            \If{$w_q^2/\varsigma_q>1+\kappa$}
                \State Update
                $\mathcal S\leftarrow \mathcal S\cup\{q\}$,
                $\boldsymbol{\Xi}_{\mathcal S}
                \leftarrow
                [\boldsymbol{\Xi}_{\mathcal S},\boldsymbol{\phi}_q]$,
                and
                $\boldsymbol{\Psi}_{\mathcal S}
                \leftarrow
                [\boldsymbol{\Psi}_{\mathcal S},\boldsymbol{\psi}_q]$.

                \State Update
                $\hat{\alpha}_q=(w_q^2-\varsigma_q)^{-1}$ and
                $\hat{\mathbf A}
                =
                {\rm diag}\{\hat{\alpha}_i:i\in\mathcal S\}$.

                \State Update $\boldsymbol{\Sigma}$, $\boldsymbol{\mu}$,
                and $\hat{\sigma}$ by \eqref{eq:45}, \eqref{eq:46},
                and \eqref{eq:48}, respectively.

                \State Reconstruct
                $\hat{\mathbf h}^{(q)}
                =
                \boldsymbol{\Psi}_{\mathcal S}\boldsymbol{\mu}$.

                \If{$
                \|\hat{\mathbf h}^{(q)}-\hat{\mathbf h}^{(q-1)}\|_2^2
                /
                \|\hat{\mathbf h}^{(q-1)}\|_2^2
                \leq \varepsilon$}
                    \State \textbf{break}
                \EndIf
            \EndIf
        \EndFor

        \State \Return \(\hat{\mathbf h}=\boldsymbol{\Psi}_{\mathcal S}\bm{\mu}\).
    \end{algorithmic}
\end{algorithm}

\section{Model Extensions and Computational Complexity Analysis}
\label{IV}
In this section, we first extend the proposed framework to multi-user pilot-contaminated and UPA-based XL-RIS models.
Then, we analyze the computational complexity of the proposed algorithms.

\subsection{Extension to Multi-User Pilot-Contamination Model}
\label{subsec:multi_user_pc}

We further extend the considered XL-RIS channel estimation framework to a multi-user uplink training scenario with pilot contamination \cite{Jose2011}. 
For RIS-assisted multi-user channel estimation, contaminated cascaded-channel observations have been considered in \cite{Yu2024KF_RIS}, where users sharing the same pilot introduce structured cascaded-channel interference. 
Motivated by this model, we adopt an energy-normalized effective pilot-contamination model to evaluate the proposed estimator under different aggregate contamination levels.
\begin{algorithm}[!t]
    \caption{Overall Channel Estimation Framework}
    \label{alg:overall_framework}
    \begin{algorithmic}[1]
        \Require \(\mathbf y\), \(\tilde{\boldsymbol{\Phi}}\), \(\mathbf D_{\rm RIS}\), \(\boldsymbol{\Gamma}\), $K$, $Q$, $\kappa$, $\epsilon$.

        \State \textbf{Offline stage: Algorithm \ref{alg:DK-ODC} \textsc{DK-ODC}}
        \State Detect the BS-side angular support
        \(\mathcal M\) from \(\mathbf y\) and \(\tilde{\boldsymbol{\Phi}}\).
        \State Refine the selected BS-side angles and update \(\boldsymbol{\Omega}\).
        \State Construct the compressed sensing dictionary \(\tilde{\boldsymbol{\Phi}}_{P}^{\downarrow}\)
        and the compressed reconstruction dictionary \(\tilde{\boldsymbol{\Psi}}_{P}^{\downarrow}\).
        \State Obtain
        \(\{
        \tilde{\boldsymbol{\Phi}}_{P}^{\downarrow},
        \tilde{\boldsymbol{\Psi}}_{P}^{\downarrow}\}\)
        for online sparse recovery.

        \State \textbf{Online stage: Algorithm \ref{alg:SI-VBL} \textsc{SI-VBL}}
        \State Screen a \(Q\)-dimensional subspace using \(\mathbf y\) and \(\tilde{\boldsymbol{\Phi}}_{P}^{\downarrow}\).
        \State Prune the candidate subspace according to \(\kappa\) and obtain the active support set \(\mathcal S\).
        \State Obtain the posterior mean \(\bm{\mu}\) until the tolerance \(\epsilon\) is satisfied.
        \State Reconstruct
        $\hat{\mathbf h}
        =
        \boldsymbol{\Psi}_{\mathcal S}\boldsymbol{\mu}$
        \State \Return \(\hat{\mathbf h}\).
    \end{algorithmic}
\end{algorithm}
Consider \(K_{\rm u}\) single-antenna UEs. 
The cascaded channel associated with the \(j\)-th UE is given by
\begin{equation}
\mathbf{G}_j
=
\mathbf{G}_{\rm RB}
\operatorname{diag}
\left(
\mathbf{h}_{{\rm UR},j}
\right),
\quad
j=1,\ldots,K_{\rm u},
\label{eq:multi_user_cascaded_channel}
\end{equation}
where \(\mathbf{G}_{\rm RB}\) is the quasi-static RIS-BS channel shared by all UEs, 
whereas \(\mathbf{h}_{{\rm UR},j}\) is user-dependent. 
For the \(j\)-th target UE, the contaminated training observation is modeled as
\begin{equation}
\mathbf{Y}^{\rm pc}_j =
\left(
\mathbf{G}_j
+
\alpha_{{\rm pc},j}
\sum_{\substack{i=1\\i\neq j}}^{K_{\rm u}}
\mathbf{G}_i
\right)\boldsymbol{\Gamma}
+
\mathbf{E}_j ,
\label{eq:pc_model_matrix}
\end{equation}
where the second term inside the parentheses represents the aggregate cascaded-channel contamination from the other UEs. 
The scaling factor \(\alpha_{{\rm pc},j}\) is chosen to impose a prescribed contamination-to-signal power ratio, i.e.,
\begin{equation}
\alpha_{{\rm pc},j}
=
\sqrt{
10^{\zeta_{\rm pc}/10}
\frac{
\left\|
\mathbf{G}_j\boldsymbol{\Gamma}
\right\|_F^2
}{
\left\|
\sum_{\substack{i=1\\i\neq j}}^{K_{\rm u}}
\mathbf{G}_i\boldsymbol{\Gamma}
\right\|_F^2
}
}.
\label{eq:pc_scaling}
\end{equation}
Accordingly, the effective pilot-contamination ratio satisfies
\begin{equation}
{\rm PCR}_j
\triangleq
\frac{
\left\|
\alpha_{{\rm pc},j}
\sum_{\substack{i=1\\i\neq j}}^{K_{\rm u}}
\mathbf{G}_i\boldsymbol{\Gamma}
\right\|_F^2
}{
\left\|
\mathbf{G}_j\boldsymbol{\Gamma}
\right\|_F^2
}
=
10^{\zeta_{\rm pc}/10},
\end{equation}
where \(\zeta_{\rm pc}\) denotes the PCR level in dB. 
A smaller \(\zeta_{\rm pc}\) corresponds to weaker pilot contamination.
For notational compactness, we define the effective contaminated cascaded channel as
\begin{equation}
\mathbf{G}^{\rm eff}_j
\triangleq
\mathbf{G}_j
+
\alpha_{{\rm pc},j}
\sum_{\substack{i=1\\i\neq j}}^{K_u}
\mathbf{G}_i .
\label{eq:effective_pc_channel}
\end{equation}
Then, \eqref{eq:pc_model_matrix} can be rewritten as
\begin{equation}
\mathbf{Y}_j^{\rm pc}
=
\mathbf{G}_{j}^{\rm eff}\bm{\Gamma}
+
\mathbf{E}_j.
\label{eq:pc_model_compact}
\end{equation}
In this evaluation, the proposed estimator takes the contaminated observation \(\mathbf{Y}^{\rm pc}_j\) as its input and estimates the desired cascaded channel of the target UE. 
The numerical evaluation of this multi-user pilot contamination model is provided in Section \ref{subsec:pilot_contamination_sim}.

\subsection{Extension to UPA-Based XL-RIS model}
\label{subsec:UPA}
The proposed framework can also be extended to a UPA-based XL-RIS. 
Consider an \(N_y\times N_z\) UPA deployed on the \(y\)-\(z\) plane, where \(N=N_yN_z\). 
The effective aperture of the UPA is characterized by
\begin{equation}
D_{\rm UPA}
=
\sqrt{\big((N_y-1)d_y\big)^2+\big((N_z-1)d_z\big)^2},
\end{equation}
and the corresponding Rayleigh distance \cite{Rayleigh} is
\begin{equation}
R_{\rm UPA}
=
\frac{2D_{\rm UPA}^2}{\lambda}.
\end{equation}
We describe the RIS-side propagation direction by the azimuth angle \(\vartheta\) and the elevation angle \(\psi\). 
Specifically, \(\vartheta\in[-\pi,\pi)\) is measured in the \(x\)-\(y\) plane with respect to the \(x\)-axis, while \(\psi\in[-\pi/2,\pi/2]\) is measured with respect to the \(x\)-\(y\) plane. 

For the far-field RIS-BS link, the normalized far-field UPA steering vector \cite{SA-LS} is written as
\begin{equation}
\mathbf{a}_{\rm UPA}(\vartheta,\psi)
=
\mathbf{a}_{z}(\psi)
\otimes
\mathbf{a}_{y}(\vartheta,\psi),
\end{equation}
where
\begin{equation}
\mathbf{a}_{y}(\vartheta,\psi)
=
\frac{1}{\sqrt{N_y}}
\left[
e^{j\frac{2\pi}{\lambda}n_yd_y\cos\psi\sin\vartheta}
\right]_{n_y=0}^{N_y-1},
\end{equation}
and
\begin{equation}
\mathbf{a}_{z}(\psi)
=
\frac{1}{\sqrt{N_z}}
\left[
e^{j\frac{2\pi}{\lambda}n_zd_z\sin\psi}
\right]_{n_z=0}^{N_z-1}.
\end{equation}

For the near-field UE-RIS link, the planar-wave approximation is no longer valid, and the spherical-wave propagation across the UPA should be explicitly characterized \cite{Liu_Low}. 
Therefore, the near-field UPA steering vector is given by
\begin{equation}
\mathbf{b}_{\rm UPA}(\vartheta,\psi,r)
=
\frac{1}{\sqrt{N}}
\left[
e^{j\frac{2\pi}{\lambda}
\left(
r(0)-r_{n_y,n_z}
\right)}
\right]_{n_y,n_z},
\end{equation}
where 
\begin{equation}
\begin{aligned}
r_{n_y,n_z}
=
\Big[
&r^2
+
(n_yd_y)^2
+
(n_zd_z)^2 \\
&-2r
\big(
n_yd_y\cos\psi\sin\vartheta
+
n_zd_z\sin\psi
\big)
\Big]^{1/2}.
\end{aligned}
\end{equation}

Substituting the RIS-BS and UE-RIS channel models yields
\begin{align}
\mathbf{G}
&=
\sqrt{\frac{MN}{L_1}}
\sqrt{\frac{N}{L_2}}
\sum_{l_1=1}^{L_1}
\sum_{l_2=1}^{L_2}
\omega_{l_1}\omega_{l_2}
\mathbf{a}_{M}
\left(
\varphi_{l_1}^{r_{\rm RB}}
\right)
\nonumber\\
&\quad
\times
\left(
\mathbf{a}_{\rm UPA}
\left(
\vartheta_{l_1}^{t_{\rm RB}},
\psi_{l_1}^{t_{\rm RB}}
\right)
\odot
\mathbf{b}_{\rm UPA}^{*}
\left(
\vartheta_{l_2}^{r_{\rm UR}},
\psi_{l_2}^{r_{\rm UR}},
r_{l_2}^{r_{\rm UR}}
\right)
\right)^H .
\label{eq:upa_cascaded_original}
\end{align}
Equation \eqref{eq:upa_cascaded_original} shows that the UPA-based cascaded channel is governed by the Hadamard product between a far-field planar-wave steering vector and a conjugated near-field spherical-wave steering vector. 
Based on the phase decoupling results for ULA-based systems in \textbf{Lemma 1}, we extend the derivation to UPA geometries and establish the corresponding equivalent structure in \textbf{Lemma 2}.

\textbf{Lemma 2:} \textit{
Let 
\(\mathbf a_{\rm UPA}
(\vartheta_{l_1}^{t_{\rm RB}},\psi_{l_1}^{t_{\rm RB}})\)
denote the far-field UPA steering vector and
\(\mathbf b_{\rm UPA}
(\vartheta_{l_2}^{r_{\rm UR}},\psi_{l_2}^{r_{\rm UR}},
r_{l_2}^{r_{\rm UR}})\)
denote the near-field UPA steering vector.
Then, their Hadamard product can be approximated as}
\begin{equation}
\begin{aligned}
&\mathbf a_{\rm UPA}
(\vartheta_{l_1}^{t_{\rm RB}},\psi_{l_1}^{t_{\rm RB}})
\odot
\mathbf b_{\rm UPA}^{*}
(\vartheta_{l_2}^{r_{\rm UR}},\psi_{l_2}^{r_{\rm UR}},
r_{l_2}^{r_{\rm UR}})  \\
&\approx
\frac{1}{N}
\left[
\exp\left(
j\frac{2\pi}{\lambda}
\Delta d_{l_1,l_2}^{\rm UPA}(n_y,n_z)
\right)
\right]_{n_y,n_z},
\label{eq:upa_hadamard}
\end{aligned}
\end{equation}
\textit{where the effective path difference at the \((n_y,n_z)\)-th RIS element is given by}
\begin{equation}
\begin{aligned}
\Delta d_{l_1,l_2}^{\rm UPA}(n_y,n_z)
&=
\underbrace{
\eta_{l_1}^{t}
-
\eta_{l_2}^{r}
}_{\text{far-field term}}
+
\underbrace{
\frac{
\rho^2-\left(\eta_{l_2}^{r}\right)^2
}{
2r_{l_2}^{r_{\rm UR}}
}
}_{\text{near-field term}},
\label{eq:upa_delta_d}
\end{aligned}
\end{equation}
\textit{with}
\begin{equation}
\begin{aligned}
\eta_{l_1}^{t}
&=
n_y d_y
\cos\psi_{l_1}^{t_{\rm RB}}
\sin\vartheta_{l_1}^{t_{\rm RB}}
+
n_z d_z
\sin\psi_{l_1}^{t_{\rm RB}}, \\
\eta_{l_2}^{r}
&=
n_y d_y
\cos\psi_{l_2}^{r_{\rm UR}}
\sin\vartheta_{l_2}^{r_{\rm UR}}
+
n_z d_z
\sin\psi_{l_2}^{r_{\rm UR}}, \\
\rho^2
&=
(n_y d_y)^2+(n_z d_z)^2 .
\end{aligned}
\label{eq:upa_eta_rho}
\end{equation}

\textit{Proof:} See Appendix D. \hfill $\blacksquare$

Based on the phase-decoupling property of \textbf{Lemma 2}, the UPA-based cascaded channel in \eqref{eq:upa_cascaded_original} can be rewritten as
\begin{equation}
\mathbf{G}
=
\sum_{l_1=1}^{L_1}
\sum_{l_2=1}^{L_2}
\beta_{l_1l_2}
\mathbf a_M
\left(
\varphi_{l_1}^{r_{\rm RB}}
\right)
\mathbf b_{\rm UPA}^{H}
\left(
\vartheta_{l_1l_2},
\psi_{l_1l_2},
r_{l_2}^{r_{\rm UR}}
\right),
\label{eq:upa_simplified_channel}
\end{equation}
where
\(
\beta_{l_1l_2}
\triangleq
\sqrt{{MN}/{L_1}}
\sqrt{{N}/{L_2}}
\omega_{l_1}\omega_{l_2}
\).
The effective parameter tuple
\(
(\vartheta_{l_1l_2},\psi_{l_1l_2},r_{l_2}^{r_{\rm UR}})
\)
is induced by the superposition of the far-field linear phase and the near-field Fresnel phase over the UPA.
Different from the ULA case, where the RIS-side sparse atom is indexed by an angle--distance pair, the UPA-based atom is indexed by an azimuth--elevation--distance tuple \cite{HSGP}. 
Hence, the RIS-side polar-domain dictionary is extended from
\(
\mathbf D_{\rm RIS}\in\mathbb C^{N\times N_\theta S}
\)
to
\begin{equation}
\mathbf D_{\rm RIS}^{\rm UPA}
=
\left[
\mathbf b_{\rm UPA}(\vartheta_i,\psi_j,r_s)
\right]_{i,j,s}
\in
\mathbb C^{N    \times N_\vartheta N_\psi S}.
\label{eq:upa_dictionary}
\end{equation}
This additional angular dimension does not alter \textbf{Algorithm \ref{alg:DK-ODC}}, since DK-ODC is used to optimize the quasi-static angular support set on the BS-side and is independent of the polar parameterization on the RIS-side.
It only enlarges the online candidate dictionary used in \textbf{Algorithm \ref{alg:SI-VBL}}, whose SI-VBL updates remain unchanged after replacing the atom index \((\theta,r)\) with \((\vartheta,\psi,r)\).
The numerical results for validating the UPA-based extension are presented in Section \ref{subsec:UPA_sim}.

\subsection{Computational Complexity Analysis}
Table~\ref{tab:complexity} summarizes the computational complexity and average running time of the considered online sparse recovery schemes.
For a fair comparison, all schemes use the same BS-side compressed dictionary
\(\widehat{\mathbf U}_{\rm BS}\) obtained by \textbf{Algorithm~\ref{alg:DK-ODC}}, where the original BS-side dimension\footnote{If a full-dimensional BS-side dictionary is used, the resulting effective sensing matrix has dimension \(MT \times M_GN_GS\), which makes the subsequent channel estimation computationally and memory intensive, and thus difficult to implement in XL-RIS systems.} is reduced from \(M_G\) to \(K\). 
Thus, the difference among the compared schemes lies in the RIS-side sparse representation and the corresponding sparse recovery method.
For the angular-domain representation, A-OMP~\cite{OMP}, A-SBL~\cite{SBL}, and A-PC-SBL~\cite{PC-SBL} are selected as benchmarks.
These methods adopt the RIS-side angular-domain dictionary 
\(\mathbf U_{\rm RIS}\in\mathbb C^{N\times N_G}\), leading to
\begin{equation}
\widetilde{\boldsymbol{\Phi}}_{\rm A}^{\downarrow}
=
\widehat{\mathbf U}_{\rm BS}
\otimes
\left(
\boldsymbol{\Gamma}^{T}
\mathbf U_{\rm RIS}
\right)
\in
\mathbb C^{MT\times KN_G}.
\label{eq:Phi_A_down}
\end{equation}
For the polar-domain representation, P-OMP~\cite{polar}, P-HSGP~\cite{HSGP}, P-LAOMP~\cite{3D-D-LAOMP}, and the proposed SI-VBL are considered.
These methods adopt the RIS-side polar-domain dictionary \(\mathbf D_{\rm RIS}\in\mathbb C^{N\times N_GS}\), leading to
\begin{equation}
\widetilde{\boldsymbol{\Phi}}_{\rm P}^{\downarrow}
=
\widehat{\mathbf U}_{\rm BS}
\otimes
\left(
\boldsymbol{\Gamma}^{T}
\mathbf D_{\rm RIS}
\right)
\in
\mathbb C^{MT\times KN_GS}.
\label{eq:Phi_P_down}
\end{equation}

\textbf{For Algorithm \ref{alg:DK-ODC}}, the computational complexity of the proposed DK-ODC is dominated by the $K$ iterations of the support detection loop.
The complexity arises from three main steps: 
1) the group support selection in \eqref{eq:23}, which entails $\mathcal{O}(KM_G N_G M T)$ for correlation operations.
2) the channel coefficient update in \eqref{eq:24}, which scales as $\mathcal{O}(M T K^2 N_G^2)$ for the LS solution. 
3) In the off-grid angle optimization in \eqref{eq:28}, the AM algorithm requires time complexity of $\mathcal{O}(Z M T K N_G + M T K^2 N_G^2)$ for $Z$ iterations. Therefore, the total complexity is approximated as $\mathcal{O}(K M_G N_G M T + Z M T K N_G + M T K^2 N_G^2)$.
\textbf{For Algorithm \ref{alg:SI-VBL}}, the complexity of the SI-VBL is primarily governed by subspace initialization and the iterative Bayesian update. 
The initial projection onto the compressed dictionary incurs a cost of $\mathcal{O}( K \cdot MT \cdot N_GS)$ in \eqref{eq:31}. 
Since the posterior covariance matrix in \eqref{eq:45} has size $R\times R$, matrix inversion costs $\mathcal{O}(R^3)$.
The total complexity can be approximated as $\mathcal{O}\!\left(K \cdot MT \cdot N_GS+R^3\right)$.
The $R^3$ inversion term is greatly reduced compared with $(KN_GS)^3$ and is usually negligible in the overall online complexity.
As shown in Table~\ref{tab:complexity}, A-OMP has the lowest average runtime because it uses the angular-domain dictionary of size \(KN_G\) and avoids distance-domain search.
However, this complexity reduction comes at the cost of severe model mismatch for the UE-RIS near-field channel, whose array response depends on both angle and distance.
Therefore, the angular-domain schemes suffer from degraded estimation accuracy in the considered hybrid-field XL-RIS scenario.
As a result, SI-VBL achieves a runtime comparable to simple polar-domain greedy algorithms, e.g., P-OMP~\cite{polar} and P-LAOMP~\cite{3D-D-LAOMP}, while providing improved estimation accuracy.
Moreover, direct polar-domain SBL or PC-SBL is computationally infeasible due to large-scale matrix inversion, and SI-VBL is still much faster than their angular-domain variants.
 \begin{table}[!t]
\centering
\caption{Per-iteration Online computational complexity and average running time comparison.}
\label{tab:complexity}
\begin{threeparttable}
    \scriptsize 
    \setlength{\tabcolsep}{6pt} 

    \begin{tabular}{lcc}
    \toprule
    \textbf{Algorithm} & \textbf{Computational Complexity $\mathcal{O}(\cdot)$} & \textbf{Avg. Runtime (s)} \\
    \midrule
    P-OMP \cite{polar}       & $\mathcal{O}(K \cdot MT \cdot N_GS)$           & 0.58 \\
    A-OMP \cite{OMP}          & $\mathcal{O}(K \cdot MT \cdot N_G)$             & 0.098 \\
    A-SBL \cite{SBL}        & $\mathcal{O}( (KN_G)^3)$ & 5.79 \\
    A-PC-SBL \cite{PC-SBL}    & $\mathcal{O}((KN_G)^3)$ & 5.57 \\
    P-HSGP \cite{HSGP} & $\mathcal{O}(K \cdot MT \cdot N_GS + I_{\rm g}MT)$ & 1.22 \\
    P-LAOMP \cite{3D-D-LAOMP} & $\mathcal{O}(K \cdot MT \cdot N_GS + MT)$ & 0.62 \\
    \textbf{Proposed SI-VBL}& $\mathcal{O}\!\left(K \cdot MT \cdot N_GS+R^3\right)$            & \textbf{0.38}\\
    \bottomrule
    \end{tabular}
    \begin{tablenotes}[flushleft]
    \scriptsize
    \item \textit{Note:} $R$ is the cardinality of the active support after pruning, satisfying $R \ll Q \ll KN_GS$.
    $I_{\rm g}$ denotes the number of gradient iterations in the HSGP algorithm. 
    \end{tablenotes}
\end{threeparttable}
\end{table}

\section{Simulation results}
\label{V}
In this section, we provide the detailed performance evaluation of the proposed algorithms for hybrid-field channel estimation. 
Unless stated otherwise in the following content, the system default configurations are all based on Table \ref{tab:II}.
\begin{table}[!t]
\centering
\caption{Simulation Configurations}
\label{tab:II}
\renewcommand{\arraystretch}{0.9}
\setlength{\tabcolsep}{9pt}
\begin{tabular}{l c c}
\toprule
\textbf{Parameter} & \textbf{Symbol} & \textbf{Value} \\
\midrule
\multicolumn{3}{c}{\textit{Common setup}} \\
\midrule
Pilot length & \(T\) & \(80\) \\
Number of paths & \(L_1,L_2\) & \(4,3\) \\
BS-side grid resolution & \(M_G\) & \(64\) \\
Number of non-zero supports & \(K\) & \(4\) \\
Candidate subspace dimension & \(Q\) & \(1000\) \\
Pruning threshold & \(\kappa\) & \(14\) \\
UE--RIS distance range & \(r_{\rm UR}\) & \([5,10]\) m \\
\midrule
\multicolumn{3}{c}{\textit{ULA-based XL-RIS setup}} \\
\midrule
Carrier frequency & \(f_c\) & \(100\) GHz \\
BS antennas, RIS elements & \(M,N\) & \(64,256\) \\
Rayleigh distance & \(R_{\rm ULA}\) & \(98.3\) m \\
RIS-side polar grid & \(N_G,S\) & \(512,6\) \\
\midrule
\multicolumn{3}{c}{\textit{UPA-based XL-RIS setup}} \\
\midrule
Carrier frequency & \(f_c\) & \(28\) GHz \\
RIS dimensions & \(N_y,N_z,N\) & \(128,4,512\) \\
Rayleigh distance & \(R_{\rm UPA}\) & \(86.5\) m \\
RIS-side polar grid & \(N_\vartheta,N_\psi,S\) & \(512,8,6\) \\
\bottomrule
\end{tabular}
\end{table}
The normalized mean squared error (NMSE) is adopted as the performance metric \cite{polar}, which is defined as
\begin{equation}
{\rm NMSE}
\triangleq
\mathbb{E}
\left[
\frac{
\left\|
\operatorname{vec}\left(\mathbf G^{T}\right)
-
\hat{\mathbf{h}}
\right\|_2^2
}{
\left\|
\operatorname{vec}\left(\mathbf G^{T}\right)
\right\|_2^2
}
\right],
\label{eq:nmse}
\end{equation}
where \(\operatorname{vec}(\mathbf G^{T})\) denotes the true vectorized cascaded channel. 
Results for each scheme are averaged over 100 Monte Carlo trials.
The AoAs and AoDs of all propagation paths are independently generated from continuous angular domains in each Monte Carlo trial.
All algorithms are implemented on a CPU i7-14650HX with 16 GB RAM.
\begin{figure}[ht]
\centering
\includegraphics[width=2.6in]{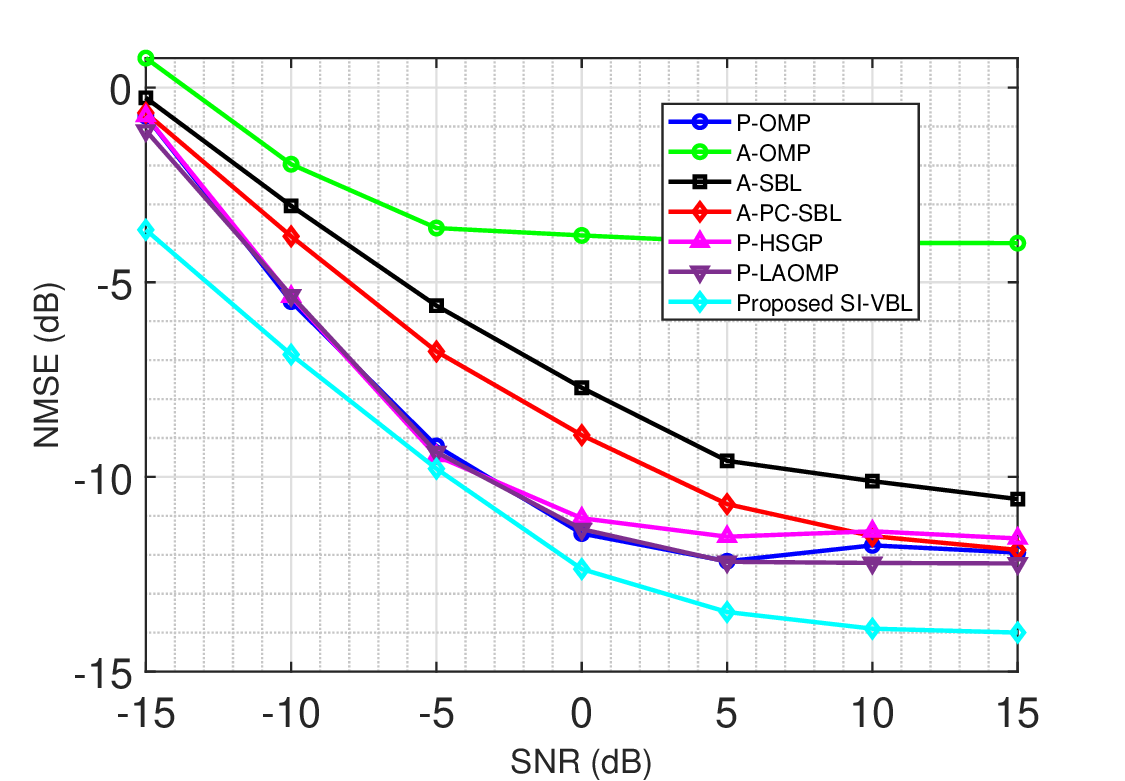}
\caption{NMSE performance comparison versus SNR in the ULA-based XL-RIS setup, with \(T=80\), \(M=64\), \(N=256\), \(L_1=4\), and \(L_2=3\).}
\label{fig:snr-ULA}
\vspace{-3mm}
\end{figure}

\begin{figure}[ht]
\centering
\includegraphics[width=2.6in]{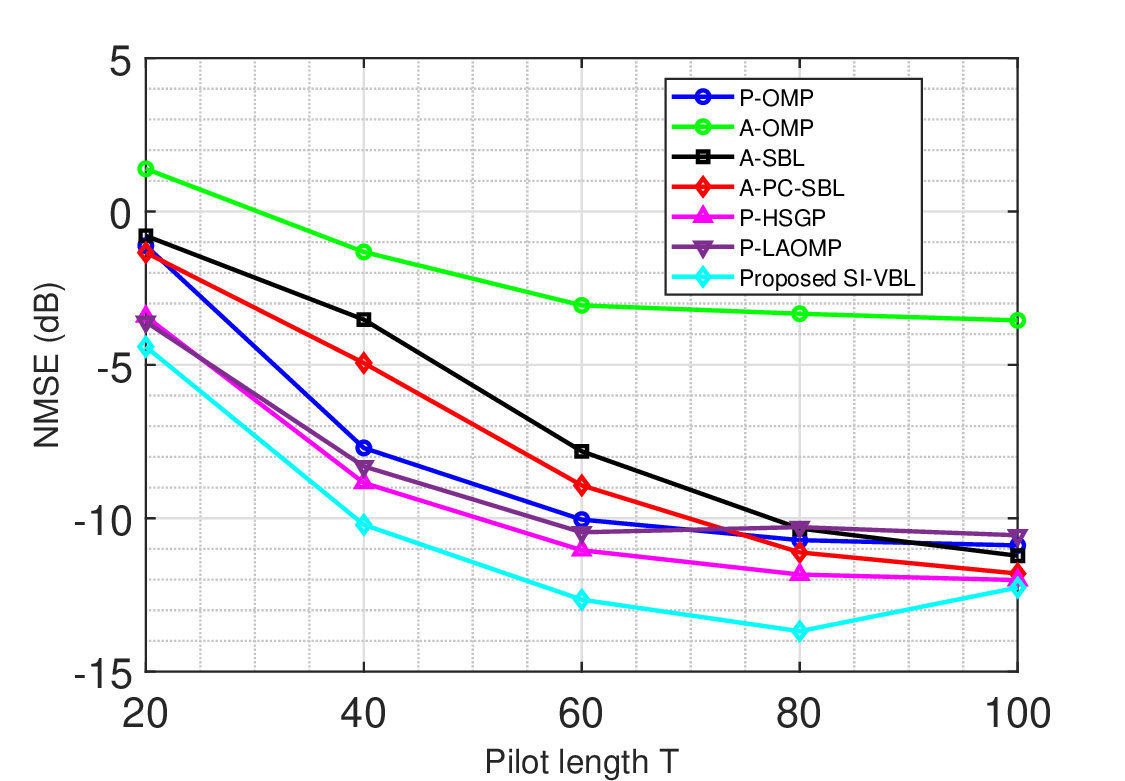}
\caption{NMSE performance comparison versus pilot length in the ULA-based XL-RIS setup, with \({\rm SNR}=5\) dB, \(M=64\), \(N=256\), \(L_1=4\), and \(L_2=3\).}
\label{fig:pilot-ULA}
\vspace{-3mm}
\end{figure}

\begin{figure}[ht]
\centering
\includegraphics[width=2.6in]{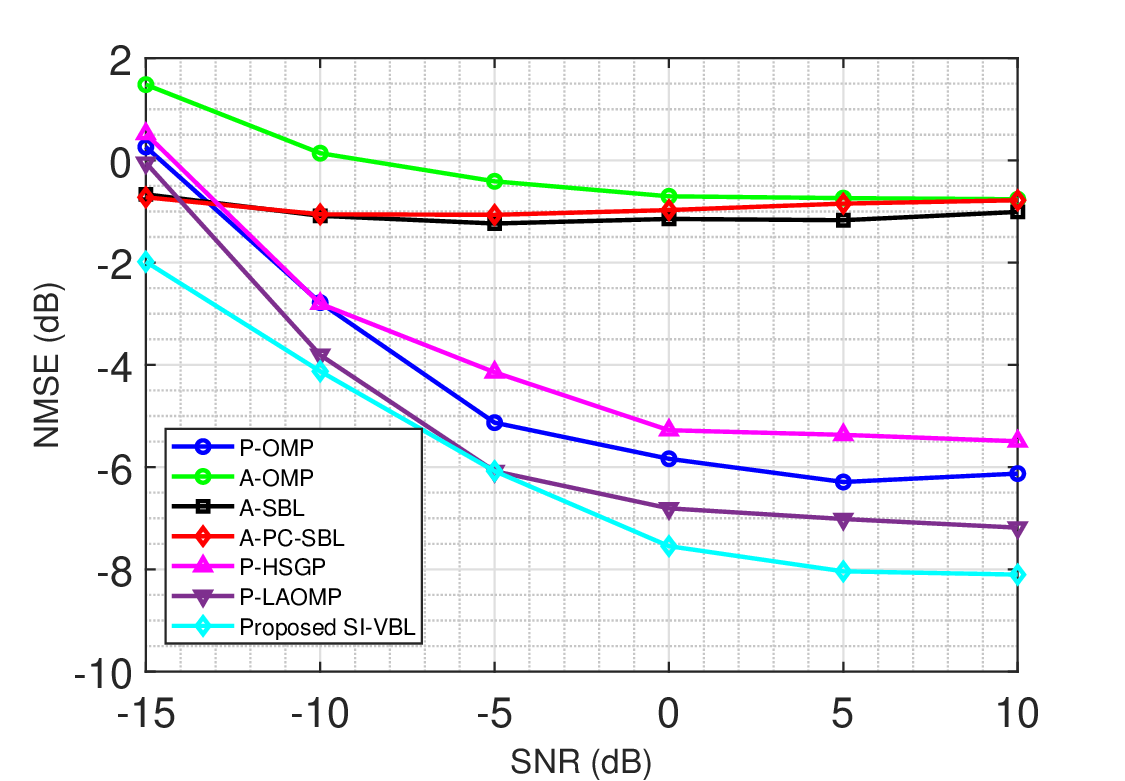}
\caption{NMSE performance comparison versus SNR in the UPA-based XL-RIS setup, with \(T=80\), \(M=64\), \(N_y=128\), \(N_z=4\), \(L_1=4\), and \(L_2=3\).}
\label{fig:snr-UPA}
\vspace{-3mm}
\end{figure}

\begin{figure}[ht]
\centering
\includegraphics[width=2.6in]{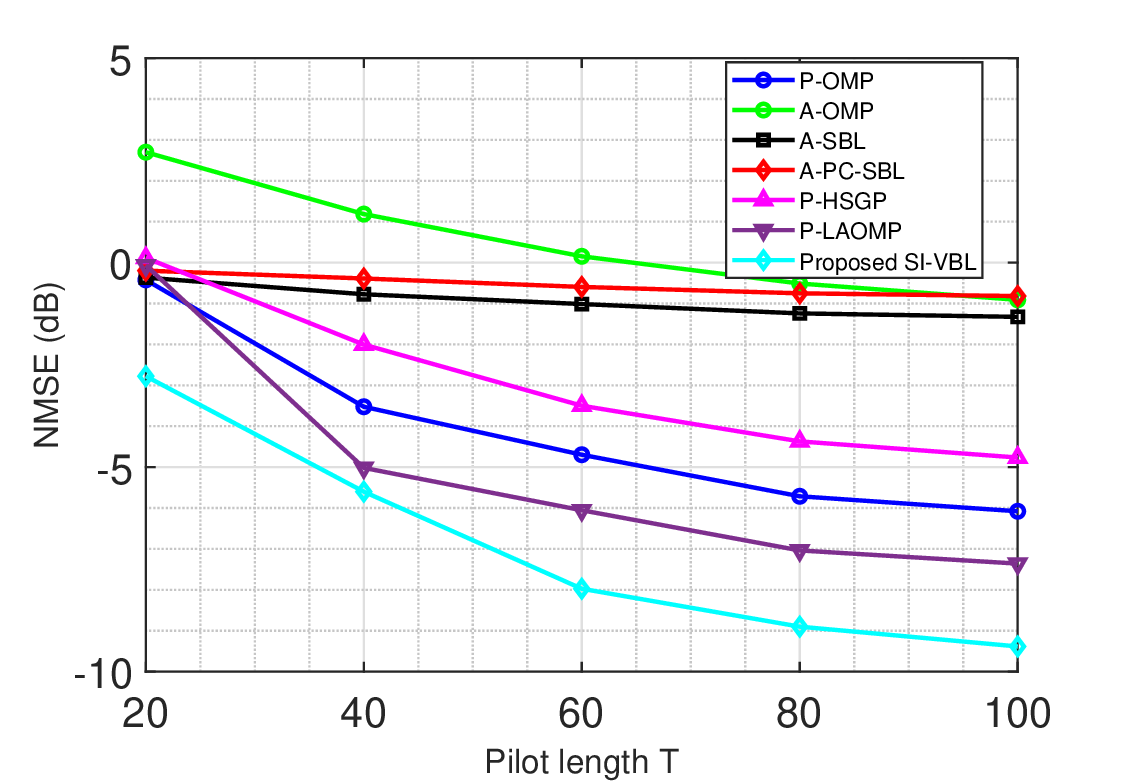}
\caption{NMSE performance comparison versus pilot length in the UPA-based XL-RIS setup, with \({\rm SNR}=5\) dB, \(M=64\), \(N_y=128\), \(N_z=4\), \(L_1=4\), and \(L_2=3\).}
\label{fig:pilot-UPA}
\vspace{-3mm}
\end{figure}

\subsection{Estimation Accuracy under ULA- and UPA-based XL-RIS}
\label{subsec:UPA_sim}
Fig.~\ref{fig:snr-ULA} and Fig.~\ref{fig:pilot-ULA} first evaluate the estimation accuracy in the ULA-based XL-RIS setup. 
As expected, increasing the SNR or the pilot length improves the NMSE performance of all schemes, since the observation becomes less noise-limited and more measurements are available for sparse recovery. However, angular-domain methods, such as A-OMP and A-SBL, show limited accuracy because they ignore the distance-dependent phase variation of the UE-RIS near-field channel. 
The polar-domain baselines provide better model matching, while the proposed SI-VBL achieves lower NMSE in most cases. 
This gain comes from the screened Bayesian recovery, where correlation-based subspace selection reduces the search dimension and likelihood-based pruning suppresses noise-dominated atoms.

Fig.~\ref{fig:snr-UPA} and Fig.~\ref{fig:pilot-UPA} further examine the extension to the UPA-based XL-RIS setup. 
Unlike the ULA case, the UPA geometry introduces azimuth--elevation--distance sampling at the RIS side, which substantially enlarges the candidate dictionary 
and increases the coherence among neighboring atoms. 
Therefore, the performance gain becomes more moderate than in the ULA case. 
Nevertheless, SI-VBL still provides the lowest NMSE over the considered SNR and pilot-length ranges, showing that the proposed subspace screening and Bayesian pruning remain effective for higher-dimensional UPA channel recovery without resorting to full-dimensional Bayesian inference.

\subsection{Parameter Sensitivity of the Proposed SI-VBL}
\begin{figure}[ht]
\centering
\includegraphics[width=2.8in]{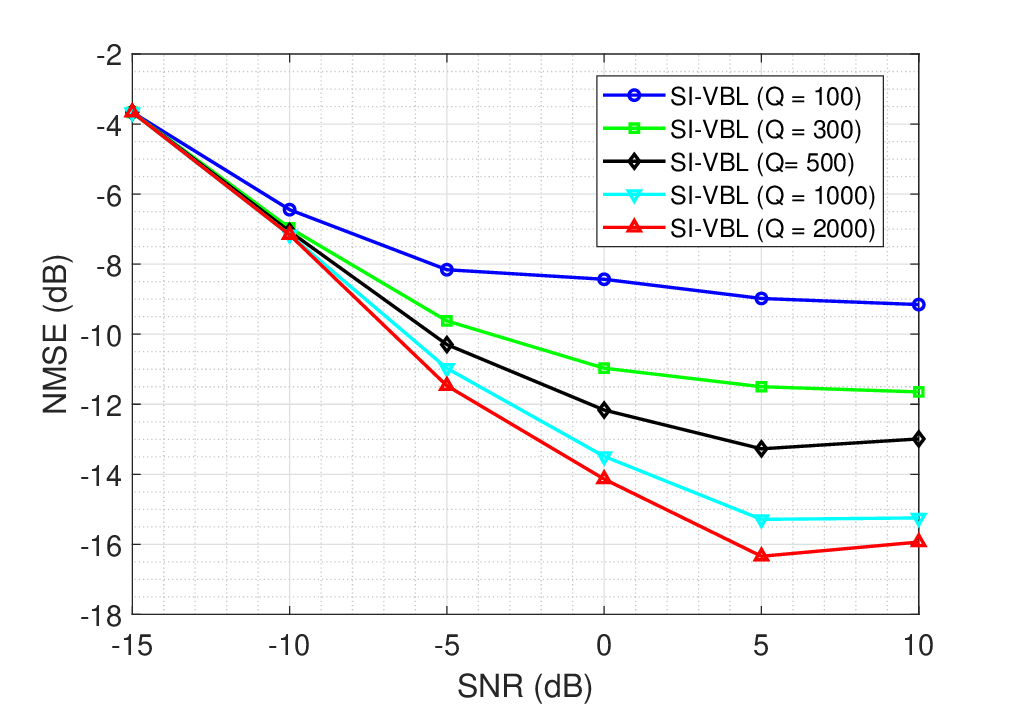}
\caption{NMSE performance of SI-VBL versus SNR under different candidate subspace dimensions \(Q\) in the ULA-based XL-RIS setup, with \(L_1=4\), \(L_2=3\), \(T=80\), and \(\kappa=14\).}
\label{diff_Q}
\end{figure}

Fig.~\ref{diff_Q} evaluates the sensitivity of SI-VBL to the retained subspace dimension \(Q\). 
As \(Q\) increases from \(100\) to \(1000\), the NMSE decreases because more candidate atoms are preserved for Bayesian refinement, whereas further increasing \(Q\) to \(2000\) yields only marginal improvement. 
This indicates an accuracy--efficiency tradeoff: a larger \(Q\) improves candidate coverage but also increases the online computational burden. 
As shown in \textbf{Algorithm~\ref{alg:SI-VBL}}, SI-VBL sequentially examines the retained atoms indexed by \(q=1,\ldots,Q\), so a larger \(Q\) generally leads to more update steps and longer runtime, which is consistent with Table~\ref{tab:runtime_Q}. 
Therefore, \(Q=1000\) is adopted as a balanced setting in the simulations.
\begin{table}[ht]
\centering
\caption{
Average Runtime of SI-VBL Under Different Candidate Subspace Dimensions}
\label{tab:runtime_Q}
\vspace{-3mm}
\renewcommand{\arraystretch}{0.9}
\setlength{\tabcolsep}{3.5pt}
\begin{tabular}{c c c c c c}
\toprule
$Q$ & $100$ & $300$ & $500$ & $1000$ & $2000$ \\
\midrule
Runtime (s) & 
$0.055$ & 
$0.091$ & 
$0.125$ & 
$0.380$ & 
$0.456$ \\
\bottomrule
\end{tabular}
\end{table}

\begin{figure}[ht]
\centering
\includegraphics[width=2.6in]{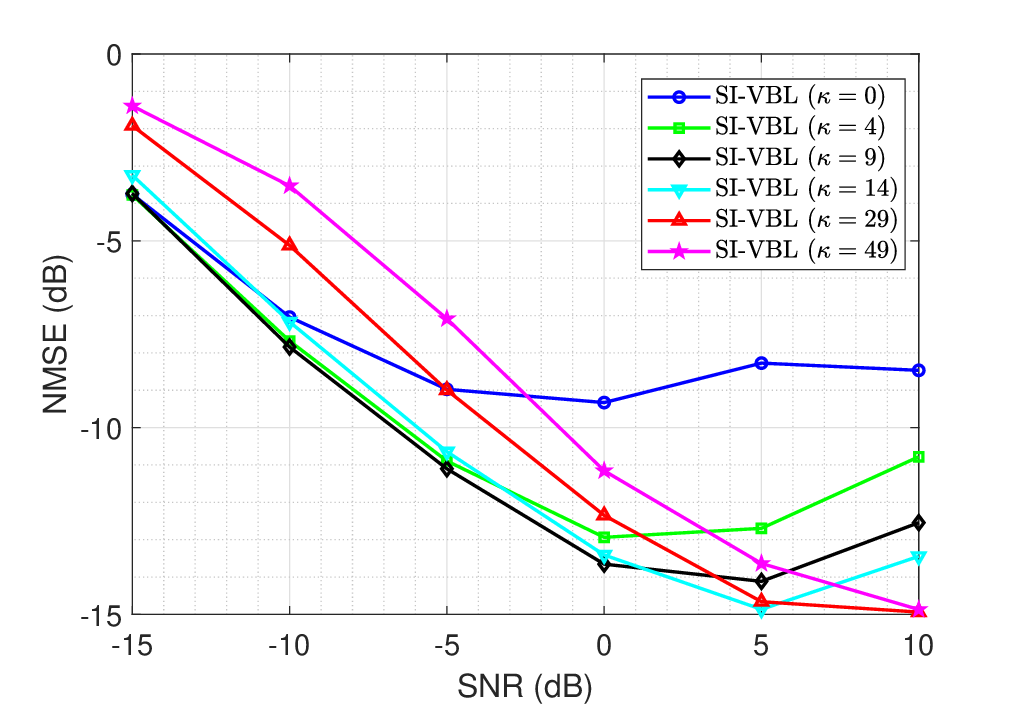}
\caption{NMSE performance of SI-VBL versus SNR under different pruning thresholds \(\kappa\) in the ULA-based XL-RIS setup, with \(L_1=4\), \(L_2=3\), \(T=80\), and \(Q=1000\).}
\label{diff_threshold}
\vspace{-3mm}
\end{figure}

Fig.~\ref{diff_threshold} shows the tradeoff controlled by \(\kappa\). 
A small \(\kappa\) corresponds to a loose admission rule, and may retain noise-induced or highly correlated false atoms in the active set. 
This can increase the posterior update dimension and degrade the recovery accuracy. 
In contrast, an overly large \(\kappa\) imposes a strict pruning rule and may remove weak but valid paths, especially in low-SNR scenarios. 
Thus, \(\kappa\) should be chosen to balance false-alarm suppression and true support preservation. 
From the simulation results in Fig.~\ref{diff_threshold}, \(\kappa=14\) achieves a favorable accuracy-robustness tradeoff and is therefore adopted as the default setting in the subsequent simulations.

\subsection{Robustness under Path-Rich Channels and Multi-User Pilot Contamination}
\label{subsec:pilot_contamination_sim}
\begin{figure}[!t]
\centering
\includegraphics[width=2.6in]{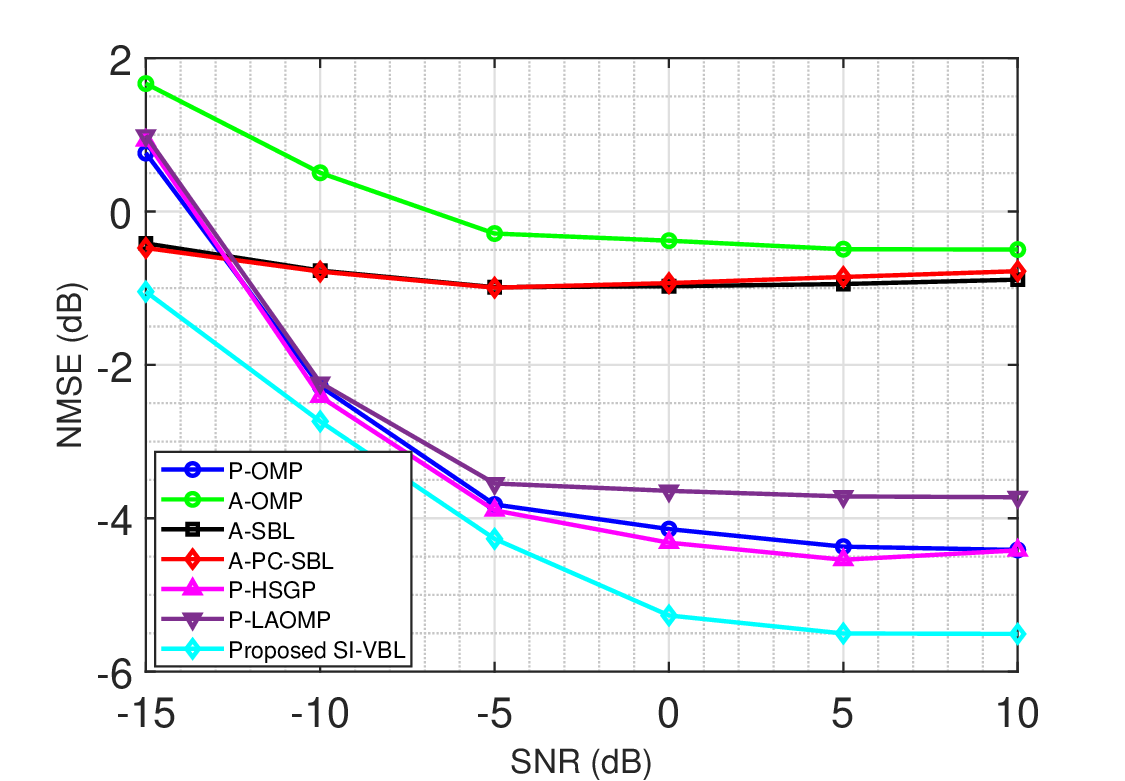}
\caption{NMSE performance versus SNR in the path-rich UPA-based XL-RIS setup, with \(L_1=L_2=5\), \(T=80\), \(\kappa=14\), and \(Q=1000\).}
\label{multi-path}
\vspace{-3mm}
\end{figure}

Fig. \ref{multi-path} considers a path-rich UPA-based XL-RIS setup.
When the number of paths increases, the active support becomes denser and the inter-path correlation becomes stronger.
Greedy algorithms are sensitive to this condition because an incorrect early support decision may propagate to later iterations. 
SI-VBL is more stable because it performs Bayesian refinement within the screened subspace and prunes unreliable atoms according to their likelihood contribution.
Therefore, although the performance gap is moderate in some low-SNR regimes, SI-VBL maintains a consistent descending NMSE trend as the SNR increases.

\begin{figure}[ht]
\centering
\includegraphics[width=2.6in]{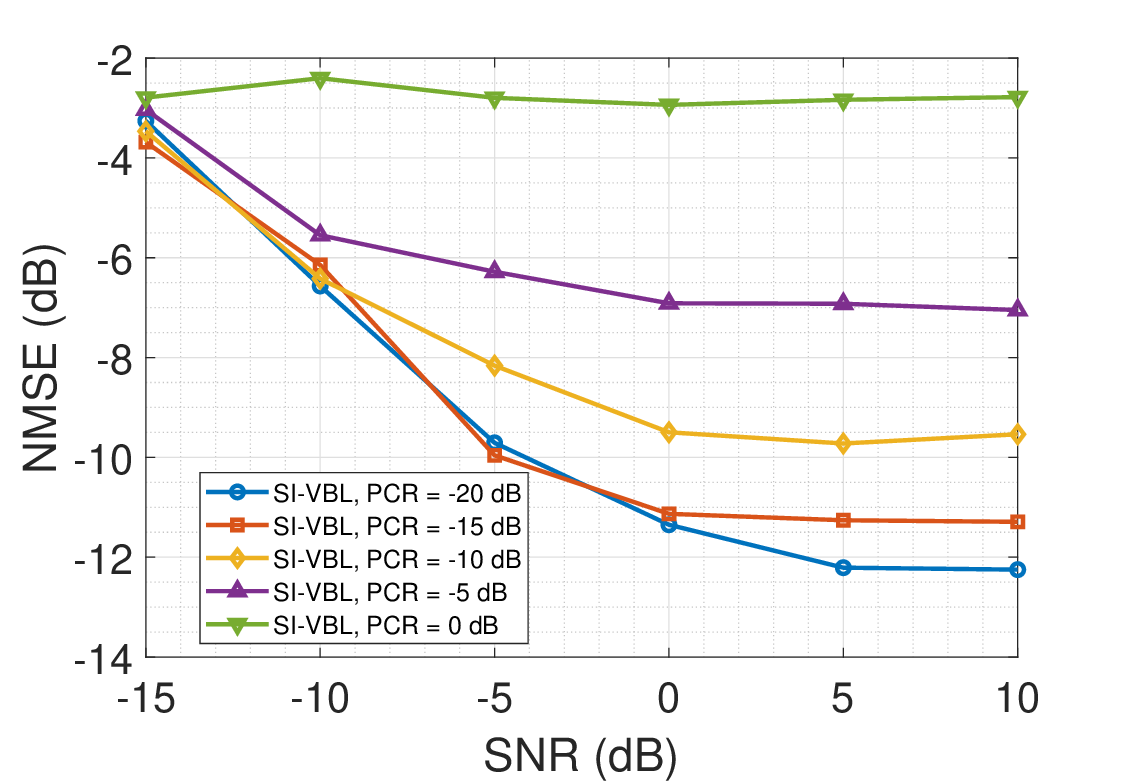}
\caption{NMSE performance of SI-VBL versus SNR under different PCR levels in the ULA-based XL-RIS setup, with \(L_1=4\), \(L_2=3\), \(T=80\), \(\kappa=14\), \(K_{\rm u}=4\), and \(Q=1000\).}
\label{diff_PCRs}
\vspace{-3mm}
\end{figure}

Fig.~\ref{diff_PCRs} evaluates the proposed SI-VBL algorithm under different PCRs in a multi-user setup with \(K_{\rm u}\) UEs, where the target UE is affected by pilot contamination from the other \(K_{\rm u}-1\) UEs.
The NMSE decreases with SNR for all PCR levels, which shows that SI-VBL can still benefit from improved observation quality.
The gap among different PCR levels becomes larger at medium and high SNRs.
This is because thermal noise is reduced in this regime, while pilot contamination remains as a structured interference term.

\begin{figure}[ht]
\centering
\includegraphics[width=2.6in]{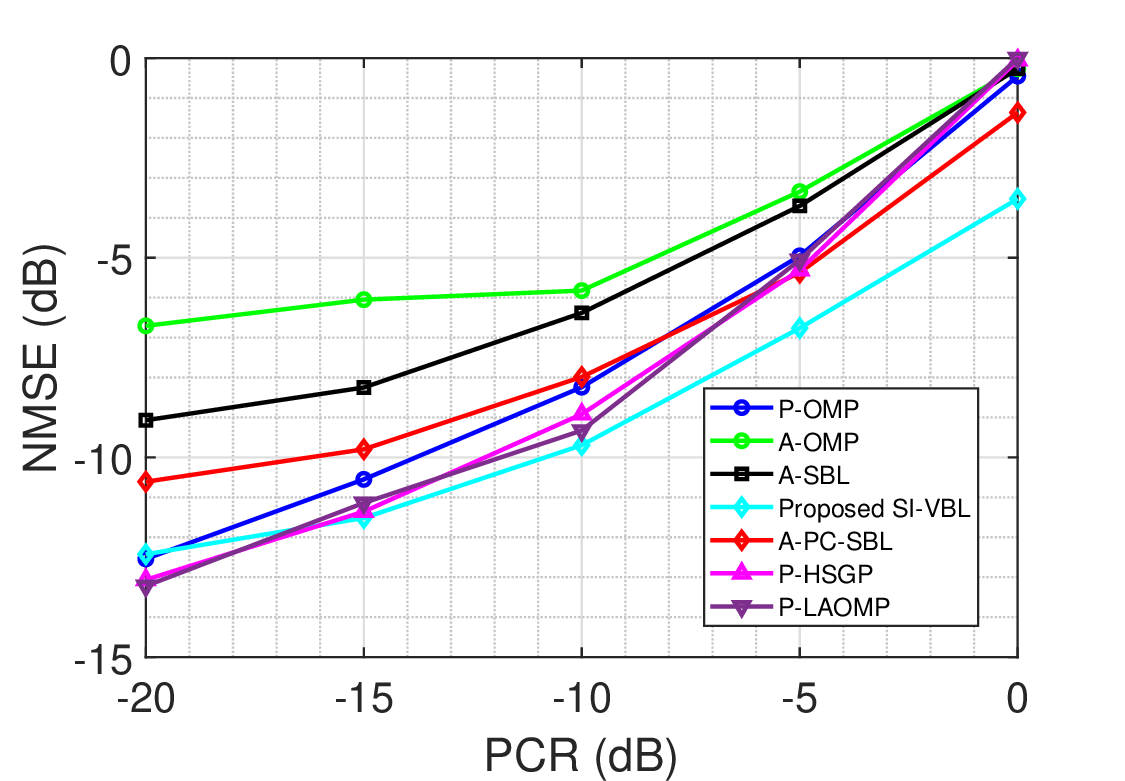}
\caption{NMSE performance comparison versus PCR for different recovery methods in the ULA-based XL-RIS setup, with \(L_1=4\), \(L_2=3\), \({\rm SNR}=5\) dB, \(T=80\), \(\kappa=14\), and \(Q=1000\).}
\label{diff_Method}
\vspace{-3mm}
\end{figure}

Fig. \ref{diff_Method} further compares different channel recovery schemes with respect to the PCR at a fixed SNR.
As expected, the NMSE of all methods increases as the PCR becomes larger, since stronger pilot contamination causes the received signal to deviate from the desired cascaded channel of the target UE. 
SI-VBL remains more robust because the screening and pruning steps help suppress contamination-induced spurious atoms while preserving dominant channel components.

\subsection{Storage Overhead and Practical Feasibility}
\label{subsec:storage}
\begin{figure}[ht]
\centering
\includegraphics[width=2.8in]{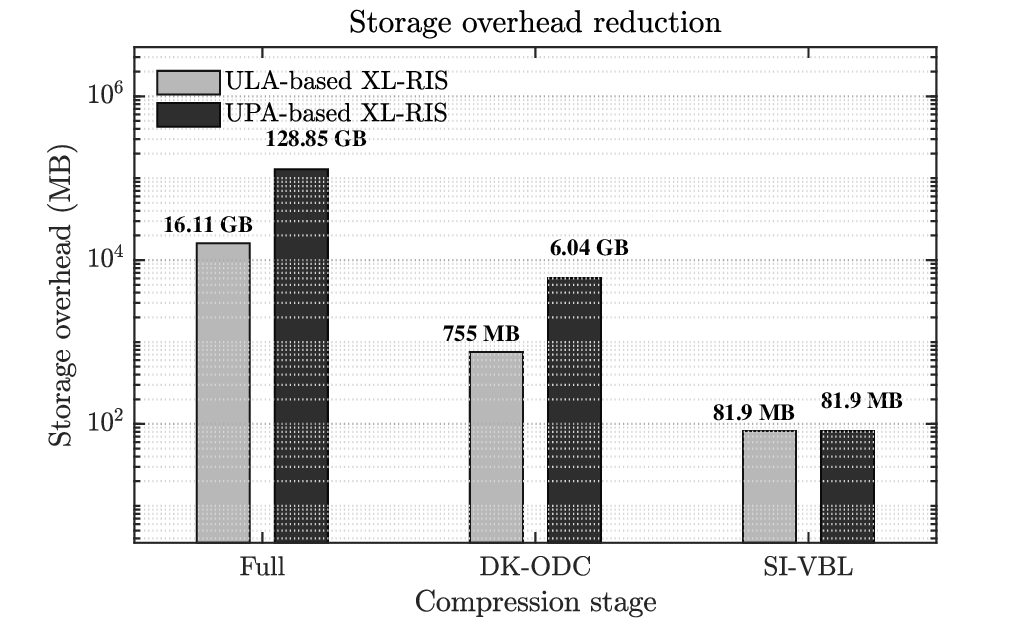}
\caption{Storage overhead comparison under ULA- and UPA-based XL-RIS configurations.}
\label{fig:storage_overhead_reduction}
\vspace{-3mm}
\end{figure}
Fig.~\ref{fig:storage_overhead_reduction} compares the storage overhead of 
the full sensing dictionary, the DK-ODC compressed dictionary, and the SI-VBL subspace dictionary. 
For a complex-valued sensing dictionary \(\bm{\Phi}\in\mathbb{C}^{MT\times N_{\rm col}}\), the storage cost is calculated as
\begin{equation}
\mathcal{S}(\bm{\Phi})=16MTN_{\rm col}\ {\rm bytes},
\label{eq:storage_formula}
\end{equation}
where each complex double-precision entry requires \(16\) bytes.

For the ULA-based XL-RIS, the full dictionary contains \(M_GN_GS\) columns, and the storage cost is
$\mathcal{S}_{\rm full}^{\rm ULA}
=
16MTM_GN_GS\ {\rm bytes}$.
After DK-ODC, the BS-side angular dimension is compressed from \(M_G\) to \(K\), leading to
$\mathcal{S}_{\rm DK}^{\rm ULA}
=
16MTKN_GS\ {\rm bytes}$.
After SI-VBL screening, online Bayesian recovery is performed over a \(Q\)-dimensional subspace, whose storage is
$
\mathcal{S}_{\rm SI}
=
16MTQ\ {\rm bytes}$.
With \(M=64\), \(T=80\), \(M_G=64\), \(N_G=512\), \(S=6\), \(K=3\), and \(Q=1000\), the ULA full dictionary requires \(16.11\) GB. 
After DK-ODC, the storage is reduced to \(755\) MB, and SI-VBL further reduces the online storage to \(81.9\) MB. 
For the UPA-based setup, with \(N_y=128\), \(N_z=4\), \(N=N_yN_z=512\), and \(N_{\vartheta}N_{\psi}S=24576\), the full dictionary requires \(128.85\) GB. 
The storage is reduced to \(6.04\) GB after DK-ODC and to \(81.9\) MB after SI-VBL screening. 
These results show that the proposed framework substantially reduces the storage burden by compressing the quasi-static BS-side angular dimension and restricting online Bayesian recovery to a low-dimensional subspace.
\section{Conclusion}
\label{VI}
This paper studied hybrid-field channel estimation for XL-RIS-assisted systems. 
Based on the double-timescale property of the cascaded channel, we propose a channel estimation framework, where the DK-ODC scheme performs BS-side off-grid dictionary compression and the SI-VBL algorithm enables low-dimensional online Bayesian recovery.
Simulation results under both ULA and UPA settings verified that the proposed framework avoids full-dimensional Bayesian recovery and achieves a favorable tradeoff among estimation accuracy, computational complexity, and storage overhead. 

Future work will extend the proposed framework toward model-driven deep learning for adaptive sparse recovery, and will consider more general XL-RIS scenarios, including multi-cell networks and multi-antenna UEs. 
Further research on hybrid-domain sparsity will be conducted, including structured sparse patterns.

\section*{Appendix A}
The $n$-th entry of the far-field steering vector is
\begin{equation}
\big[\mathbf a_N(\phi_{l_1}^{t_{\rm RB}})\big]_n
=\frac{1}{\sqrt N}\exp\!\left(j\frac{2\pi}{\lambda}nd\cos\phi_{l_1}^{t_{\rm RB}}\right),
\end{equation}
where $\forall \textit{n}\in \left\{ 0,1,\cdots ,N-1 \right\}$. The near-field steering vector can be written as
\begin{equation}
\big[\mathbf b_N^{*}(\phi_{l_2}^{r_{\rm UR}},r_{l_2}^{r_{\rm UR}})\big]_n
=\frac{1}{\sqrt N}\exp\!\left(-j\frac{2\pi}{\lambda}\big(r^{(0)}-r^{(n)}\big)\right).
\end{equation}
Therefore, the $n$-th entry of the Hadamard product is given by
\begin{equation}
\begin{aligned}
\big[\mathbf a_N(\phi_{l_1}^{t_{\rm RB}})\odot \mathbf b_N^{*}(\phi_{l_2}^{r_{\rm UR}},r_{l_2}^{r_{\rm UR}})\big]_n
&=\frac{1}{N}\exp\!\Big(j\frac{2\pi}{\lambda}\Big(
nd\cos\phi_{l_1}^{t_{\rm RB}} \\
& -\big(r^{(0)}-r^{(n)}\big)
\Big)\Big).
\end{aligned}
\end{equation}
Under the second-order approximation in the radiating near-field region, we have
\begin{equation}
r^{(0)}-r^{(n)} \approx nd\cos\phi_{l_2}^{r_{\rm UR}}-\frac{n^2d^2\sin^2\phi_{l_2}^{r_{\rm UR}}}{2r_{l_2}^{r_{\rm UR}}}.
\end{equation}
Substituting the above into the Hadamard product yields
\begin{equation}
\big[\mathbf a_N(\phi_{l_1}^{t_{\rm RB}})\odot \mathbf b_N^{*}(\phi_{l_2}^{r_{\rm UR}},r_{l_2}^{r_{\rm UR}})\big]_n
\approx\frac{1}{N}\exp\!\left(j\frac{2\pi}{\lambda}\Delta d''(n)\right),
\end{equation}
where
\begin{equation}
\Delta d''(n)=nd\big(\cos\phi_{l_1}^{t_{\rm RB}}-\cos\phi_{l_2}^{r_{\rm UR}}\big)
+\frac{n^2d^2\sin^2\phi_{l_2}^{r_{\rm UR}}}{2r_{l_2}^{r_{\rm UR}}}.
\end{equation}

\section*{Appendix B}
The $\hat{m}$-th entry of the vector $\mathbf{z}\in {{\mathbb{C}}^{{{M}_{G}}\times 1}}$ is expressed as:
\begin{equation}
\begin{aligned}
 {{z}_{\hat{m}}} &= {{\mathbf{a}}_{M}}{{({{\varphi }^{\hat{m}}})}^{\mathrm{H}}}{{\mathbf{a}}_{M}}(\varphi _{{{l}_{1}}}^{{{r}_{\mathrm{RB}}}}) \\
&= \sum_{m=0}^{M-1} \left[ \mathbf{a}_M(\varphi^{(\hat{m})}) \right]_m^* \left[ \mathbf{a}_M(\varphi_{l_1}^{r_{\mathrm{RB}}}) \right]_m.
\label{eq:59}
\end{aligned}
\end{equation}
Substituting the explicit form of the steering vector $\mathbf{a}_M(\cdot)$ into \eqref{eq:59}: 
\begin{equation}
\begin{aligned}
z_{\hat{m}} &= \frac{1}{M} \sum_{m=0}^{M-1} e^{-j \frac{2\pi d}{\lambda} m \cos(\varphi^{(\hat{m})})} \cdot e^{j \frac{2\pi d}{\lambda} m \cos(\varphi_{l_1}^{r_{\mathrm{RB}}})} \\
&= \frac{1}{M} \sum_{m=0}^{M-1} \exp\left( -j \frac{2\pi d}{\lambda} m \Delta_{\hat{m}} \right),
\label{eq:60}
\end{aligned}
\end{equation}
where $\Delta_{\hat{m}} \triangleq \cos(\varphi^{(\hat{m})}) - \cos(\varphi_{l_1}^{r_{\mathrm{RB}}})$ denotes the directional cosine difference. 
Recognizing \eqref{eq:60} as a finite geometric series sum $\sum_{n=0}^{M-1} q^n = \frac{1-q^M}{1-q}$ with $q = e^{-j \frac{2\pi d}{\lambda} \Delta_{\hat{m}}}$, we obtain
\begin{equation}
z_{\hat{m}} = \frac{1}{M} \cdot \frac{1 - e^{-j \frac{2\pi d}{\lambda} M \Delta_{\hat{m}}}}{1 - e^{-j \frac{2\pi d}{\lambda} \Delta_{\hat{m}}}}.
\label{eq:61}
\end{equation}
By extracting the half-angle phase factors from both the numerator and the denominator, i.e., applying the identity $1 - e^{-j\theta} = e^{-j\frac{\theta}{2}} \cdot 2j\sin(\frac{\theta}{2})$, \eqref{eq:61} is reformulated as
\begin{equation}
\begin{aligned}
z_{\hat{m}} &= \frac{1}{M} \frac{e^{-j \frac{\pi d}{\lambda} M \Delta_{\hat{m}}} \cdot 2j \sin \left( \frac{\pi M d}{\lambda} \Delta_{\hat{m}} \right)}{e^{-j \frac{\pi d}{\lambda} \Delta_{\hat{m}}} \cdot 2j \sin \left( \frac{\pi d}{\lambda} \Delta_{\hat{m}} \right)} \\
&= \frac{1}{M} e^{-j \frac{\pi d}{\lambda} (M-1) \Delta_{\hat{m}}} \frac{\sin \left( \frac{\pi M d}{\lambda} \Delta_{\hat{m}} \right)}{\sin \left( \frac{\pi d}{\lambda} \Delta_{\hat{m}} \right)}.
\end{aligned}
\end{equation}
Defining the Dirichlet kernel function $D_M(\cdot)$, we obtain
\begin{equation}
z_{\hat{m}} = \frac{1}{M} e^{-j \frac{\pi (M-1) d}{\lambda} \Delta_{\hat{m}}} D_M(\Delta_{\hat{m}}).
\end{equation}

\section*{Appendix C}
\subsection{Update of Channel Coefficients \texorpdfstring{$\mathbf{g}_{[\hat{m}]}$}{g\_[m\^{}hat]}}

For a fixed angular parameter $\tilde{\Theta}^{(k)}$,
the optimization with respect to $\mathbf{g}_{[\hat{m}]}$
reduces to a linear LS problem:
\begin{equation}
\mathbf{g}_{[\hat{m}]}^{(k)}
=
\underset{\mathbf{g}}{\arg\min}
\;
\left\|
\mathbf{y}
-
\widetilde{\mathbf{\Phi}}_{[\hat{m}]}\!\left(\tilde{\Theta}^{(k)}\right)
\mathbf{g}
\right\|_{2}^{2}.
\end{equation}
The closed-form solution is given by
\begin{equation}
\mathbf{g}_{[\hat{m}]}^{(k)}
=
\left(
\widetilde{\mathbf{\Phi}}_{[\hat{m}]}\!\left(\tilde{\Theta}^{(k)}\right)
\right)^{\dagger}
\mathbf{y},
\end{equation}
\subsection{Compute the gradient of the angular parameter \texorpdfstring{$\tilde{\Theta}^{(k)}$}{Θ\~(k)}}

With $\mathbf{g}_{[\hat{m}]}^{(k)}$ fixed, the objective function becomes
\begin{equation}
\mathcal{J}(\tilde{\Theta})
=
\left\|
\mathbf{r}^{(k)}(\tilde{\Theta})
\right\|_{F}^{2},
\quad
\mathbf{r}^{(k)}(\tilde{\Theta})
\triangleq
\mathbf{y}
-
\widetilde{\mathbf{\Phi}}_{[\hat{m}]}(\tilde{\Theta})
\mathbf{g}_{[\hat{m}]}^{(k)} .
\end{equation}
Using the identity
$\| \mathbf{A} \|_{F}^{2} = \mathrm{Tr}(\mathbf{A}^{H}\mathbf{A})$,
the gradient of $\mathcal{J}(\tilde{\Theta})$ with respect to the real-valued
parameter $\tilde{\Theta}$ is given by
\begin{equation}
\label{eq:theta_gradient_general}
\frac{\partial \mathcal{J}}{\partial \tilde{\Theta}}
=
2\,\mathrm{Re}
\left\{
\mathrm{Tr}
\left(
\left(
\frac{\partial \mathbf{r}^{(k)}(\tilde{\Theta})}{\partial \tilde{\Theta}}
\right)^{H}
\mathbf{r}^{(k)}(\tilde{\Theta})
\right)
\right\}.
\end{equation}

Since $\mathbf{Y}$ is independent of $\tilde{\Theta}$,
the derivative of the residual is
\begin{equation}
\frac{\partial \mathbf{r}^{(k)}(\tilde{\Theta})}{\partial \tilde{\Theta}}
=
-
\frac{\partial \widetilde{\mathbf{\Phi}}_{[\hat{m}]}(\tilde{\Theta})}
{\partial \tilde{\Theta}}
\mathbf{g}_{[\hat{m}]}^{(k)} .
\end{equation}
Substituting this expression into \eqref{eq:theta_gradient_general}
yields
\begin{equation}
\label{eq:theta_gradient_expanded}
\frac{\partial \mathcal{J}}{\partial \tilde{\Theta}}
=
-2\,\mathrm{Re}
\left\{
\mathrm{Tr}
\left(
\left[
\frac{\partial \widetilde{\mathbf{\Phi}}_{[\hat{m}]}(\tilde{\Theta})}
{\partial \tilde{\Theta}}
\mathbf{g}_{[\hat{m}]}^{(k)}
\right]^{H}
\mathbf{r}^{(k)}(\tilde{\Theta})
\right)
\right\}.
\end{equation}

The sub-dictionary $\widetilde{\mathbf{\Phi}}_{[\hat{m}]}(\tilde{\Theta})$
contains a single column that depends on $\tilde{\Theta}$, which can be written as
\begin{equation}
\widetilde{\mathbf{\Phi}}_{[\hat{m}]}(\tilde{\Theta})
=
\mathbf{u}_{M}(\tilde{\Theta})
\otimes
(\mathbf{\Gamma}^{T}\mathbf{U}_{\mathrm{RIS}}).
\end{equation}
Therefore,
\begin{equation}
\frac{\partial \widetilde{\mathbf{\Phi}}_{[\hat{m}]}(\tilde{\Theta})}
{\partial \tilde{\Theta}}
=
\frac{\partial \mathbf{u}_{M}(\tilde{\Theta})}{\partial \tilde{\Theta}}
\otimes
(\mathbf{\Gamma}^{T}\mathbf{U}_{\mathrm{RIS}}).
\end{equation}
For a ULA with normalized spatial frequency $\tilde{\Theta}$,
the array response is
\begin{equation}
\mathbf{u}_{M}(\tilde{\Theta})
=
\frac{1}{\sqrt{M}}
\left[
1,
e^{-j\pi \tilde{\Theta}},
\ldots,
e^{-j\pi (M-1)\tilde{\Theta}}
\right]^{T},
\end{equation}
whose derivative is given by
\begin{equation}
\frac{\partial \mathbf{u}_{M}(\tilde{\Theta})}{\partial \tilde{\Theta}}
=
-\,j\pi
\left(
\mathbf{m}
\odot
\mathbf{u}_{M}(\tilde{\Theta})
\right),
\quad
\mathbf{m} = [0,1,\ldots,M-1]^{T}.
\end{equation}

Substituting the above results into \eqref{eq:theta_gradient_expanded},
the gradient of the objective function with respect to $\tilde{\Theta}$
is finally obtained as
\begin{equation}
\begin{aligned}
\frac{\partial \mathcal{J}}{\partial \tilde{\Theta}}
&=
-2\,\mathrm{Re}
\Bigg\{
\mathrm{Tr}
\Bigg(
\Big[
\big(
-\,j\pi
(\mathbf{m}\odot\mathbf{u}_{M}(\tilde{\Theta}))
\otimes
(\mathbf{\Gamma}^{T}\mathbf{U}_{\mathrm{RIS}})
\big)\\
& \times
\mathbf{g}_{[\hat{m}]}^{(k)}
\Big]^{H}
\mathbf{r}^{(k)}(\tilde{\Theta})
\Bigg)
\Bigg\}.
\end{aligned}
\end{equation}
The refined angle $\tilde{\Theta}^{(k+1)}$ is obtained by performing a
projected line search along the negative gradient direction within
the region $\mathcal{I}^{(k)} = [\Theta_d, \Theta_u]$.

\section*{Appendix D}

For the UPA geometry, the \((n_y,n_z)\)-th entry of the
far-field RIS-side steering vector can be written as
\begin{equation}
\left[
\mathbf a_{\rm UPA}
\left(
\vartheta_{l_1}^{t_{\rm RB}},
\psi_{l_1}^{t_{\rm RB}}
\right)
\right]_{n_y,n_z}
=
\frac{1}{\sqrt N}
\exp\left(
j\frac{2\pi}{\lambda}
\eta_{l_1}^{t}
\right),
\end{equation}
where
\begin{equation}
\eta_{l_1}^{t}
=
n_y d_y
\cos\psi_{l_1}^{t_{\rm RB}}
\sin\vartheta_{l_1}^{t_{\rm RB}}
+
n_z d_z
\sin\psi_{l_1}^{t_{\rm RB}} .
\end{equation}
Similarly, the conjugated near-field UPA steering vector is given by
\begin{equation}
\begin{aligned}
&\big[
\mathbf b_{\rm UPA}^{*}
\big(
\vartheta_{l_2}^{r_{\rm UR}},
\psi_{l_2}^{r_{\rm UR}},
r_{l_2}^{r_{\rm UR}}
\big)
\big]_{n_y,n_z}  \\
&\quad =
\frac{1}{\sqrt N}
\exp\left[
j\frac{2\pi}{\lambda}
\big(
r_{n_y,n_z}-r_{l_2}^{r_{\rm UR}}
\big)
\right].
\end{aligned}
\end{equation}
Therefore, the element-wise Hadamard product becomes
\begin{equation}
\begin{aligned}
&\left[
\mathbf a_{\rm UPA}
\left(
\vartheta_{l_1}^{t_{\rm RB}},
\psi_{l_1}^{t_{\rm RB}}
\right)
\odot
\mathbf b_{\rm UPA}^{*}
\left(
\vartheta_{l_2}^{r_{\rm UR}},
\psi_{l_2}^{r_{\rm UR}},
r_{l_2}^{r_{\rm UR}}
\right)
\right]_{n_y,n_z} \\
&=
\frac{1}{N}
\exp\left(
j\frac{2\pi}{\lambda}
\left[
\eta_{l_1}^{t}
+
r_{n_y,n_z}
-
r_{l_2}^{r_{\rm UR}}
\right]
\right).
\end{aligned}
\end{equation}

Next, define
\begin{equation}
\eta_{l_2}^{r}
=
n_y d_y
\cos\psi_{l_2}^{r_{\rm UR}}
\sin\vartheta_{l_2}^{r_{\rm UR}}
+
n_z d_z
\sin\psi_{l_2}^{r_{\rm UR}},
\end{equation}
and
\begin{equation}
\rho^2=(n_y d_y)^2+(n_z d_z)^2 .
\end{equation}
Then, the spherical propagation distance can be expressed as
\begin{equation}
r_{n_y,n_z}
=
\left[
\left(r_{l_2}^{r_{\rm UR}}\right)^2
+\rho^2
-
2r_{l_2}^{r_{\rm UR}}\eta_{l_2}^{r}
\right]^{1/2}.
\end{equation}
Equivalently,
\begin{equation}
r_{n_y,n_z}
=
r_{l_2}^{r_{\rm UR}}
\left[
1
-
\frac{2\eta_{l_2}^{r}}{r_{l_2}^{r_{\rm UR}}}
+
\frac{\rho^2}{\left(r_{l_2}^{r_{\rm UR}}\right)^2}
\right]^{1/2}.
\end{equation}
Using the second-order Taylor expansion
\((1+x)^{1/2}\approx 1+x/2-x^2/8\), and neglecting
third and higher-order terms, we obtain
\begin{equation}
r_{n_y,n_z}
\approx
r_{l_2}^{r_{\rm UR}}
-
\eta_{l_2}^{r}
+
\frac{
\rho^2-\left(\eta_{l_2}^{r}\right)^2
}{
2r_{l_2}^{r_{\rm UR}}
}.
\end{equation}
Thus,
\begin{equation}
r_{n_y,n_z}
-
r_{l_2}^{r_{\rm UR}}
\approx
-
\eta_{l_2}^{r}
+
\frac{
\rho^2-\left(\eta_{l_2}^{r}\right)^2
}{
2r_{l_2}^{r_{\rm UR}}
}.
\end{equation}
Substituting this approximation into the Hadamard product yields
\begin{equation}
\begin{aligned}
&\left[
\mathbf a_{\rm UPA}
\left(
\vartheta_{l_1}^{t_{\rm RB}},
\psi_{l_1}^{t_{\rm RB}}
\right)
\odot
\mathbf b_{\rm UPA}^{*}
\left(
\vartheta_{l_2}^{r_{\rm UR}},
\psi_{l_2}^{r_{\rm UR}},
r_{l_2}^{r_{\rm UR}}
\right)
\right]_{n_y,n_z} \\
&\approx
\frac{1}{N}
\exp\left(
j\frac{2\pi}{\lambda}
\left[
\eta_{l_1}^{t}
-
\eta_{l_2}^{r}
+
\frac{
\rho^2-\left(\eta_{l_2}^{r}\right)^2
}{
2r_{l_2}^{r_{\rm UR}}
}
\right]
\right).
\end{aligned}
\end{equation}
By defining
\begin{equation}
\Delta d_{l_1,l_2}^{\rm UPA}(n_y,n_z)
=
\eta_{l_1}^{t}
-
\eta_{l_2}^{r}
+
\frac{
\rho^2-\left(\eta_{l_2}^{r}\right)^2
}{
2r_{l_2}^{r_{\rm UR}}
}.
\end{equation}

\end{document}